\documentclass[journal]{IEEEtran}

\usepackage{colortbl,booktabs}
\usepackage{amsmath}
\usepackage{times}
\usepackage{epsfig}
\usepackage{graphicx}
\usepackage{amsmath}
\usepackage{amssymb}
\usepackage{color}
\usepackage{wrapfig}
\usepackage{multirow}
\usepackage{cite}
\usepackage{algorithm}
\usepackage{algorithmic}
\usepackage{enumerate}

\ifCLASSINFOpdf
\else
\fi
\begin{document}

\title{UPHDR-GAN: Generative Adversarial Network for High Dynamic Range Imaging with Unpaired Data}

\author{Ru~Li,~\IEEEmembership{Student Member,~IEEE,}
        Chuan~Wang,
        Jue~Wang,~\IEEEmembership{Senior Member,~IEEE,}
        Guanghui~Liu,~\IEEEmembership{Senior Member,~IEEE,}
        Heng-Yu Zhang,
        Bing~Zeng,~\IEEEmembership{Fellow,~IEEE,}
        Shuaicheng~Liu,~\IEEEmembership{Member,~IEEE}

\thanks{
Manuscript received December 31, 2021; revised March 16, 2022, May 31, 2022 and June 22, 2022; accepted June 30, 2022.
This work was supported by the National Natural Science Foundation of China (NSFC) under Grant 62071097, Grant 61872067, Grant 62031009, and Grant 61720106004. (\emph{Corresponding authors: Guanghui Liu; Heng-Yu Zhang}.)
}

\thanks{Ru~Li, Guanghui~Liu, Bing~Zeng and Shuaicheng~Liu are with School of Information and Communication Engineering, University of Electronic Science and Technology of China, Chengdu 611731, China.  (e-mail: guanghuiliu@uestc.edu.cn)

Chuan Wang is with Megvii Technology, Chengdu 610095, China.

Jue Wang is with Tencent AI Lab, Shenzhen 518000, China.

Heng-Yu Zhang is with the Department of Cardiology, West China Hospital, Sichuan University, Chengdu 610041, China. (e-mail: zhanghengyu@wchscu.cn)

Digital Object Identifier 10.1109/TCSVT.2022.3190057
% Copyright \textcopyright 2022 IEEE. Personal use of this material is permitted. However, permission to use this material for any other purposes must be obtained from the IEEE by sending an email to pubs-permissions@ieee.org.
}
}


\maketitle

% As a general rule, do not put math, special symbols or citations in the abstract or keywords.
\begin{abstract}
The paper proposes a method to effectively fuse multi-exposure inputs and generate high-quality high dynamic range (HDR) images with unpaired datasets. Deep learning-based HDR image generation methods rely heavily on paired datasets. The ground truth images play a leading role in generating reasonable HDR images. Datasets without ground truth are hard to be applied to train deep neural networks. Recently, Generative Adversarial Networks (GAN) have demonstrated their potentials of translating images from source domain $X$ to target domain $Y$ in the absence of paired examples. In this paper, we propose a GAN-based network for solving such problems while generating enjoyable HDR results, named UPHDR-GAN. The proposed method relaxes the constraint of the paired dataset and learns the mapping from the LDR domain to the HDR domain. Although the pair data are missing, UPHDR-GAN can properly handle the ghosting artifacts caused by moving objects or misalignments with the help of the modified GAN loss, the improved discriminator network and the useful initialization phase. The proposed method preserves the details of important regions and improves the total image perceptual quality. Qualitative and quantitative comparisons against the representative methods demonstrate the superiority of the proposed UPHDR-GAN.
\end{abstract}


% \fi
\IEEEpeerreviewmaketitle

\section{Introduction}

\IEEEPARstart{T}{he} dynamic range of commercial imaging products is lower than natural scenes. 
Most digital photography sensors cannot acquire the irradiance range that is wide enough.
High dynamic range (HDR) imaging techniques have been introduced because they can overcome such limitations and generate images with a wider dynamic range. The specialized hardware device~\cite{Tumblin2005Why} 
has been introduced to directly obtain HDR images, but it is usually too expensive to be widely adopted. An optional strategy is to merge a stack of images with different exposures to produce an informative output~\cite{li2019hybrid}.

\begin{figure}[t]
   \centering
   \includegraphics[width=1.0\linewidth]{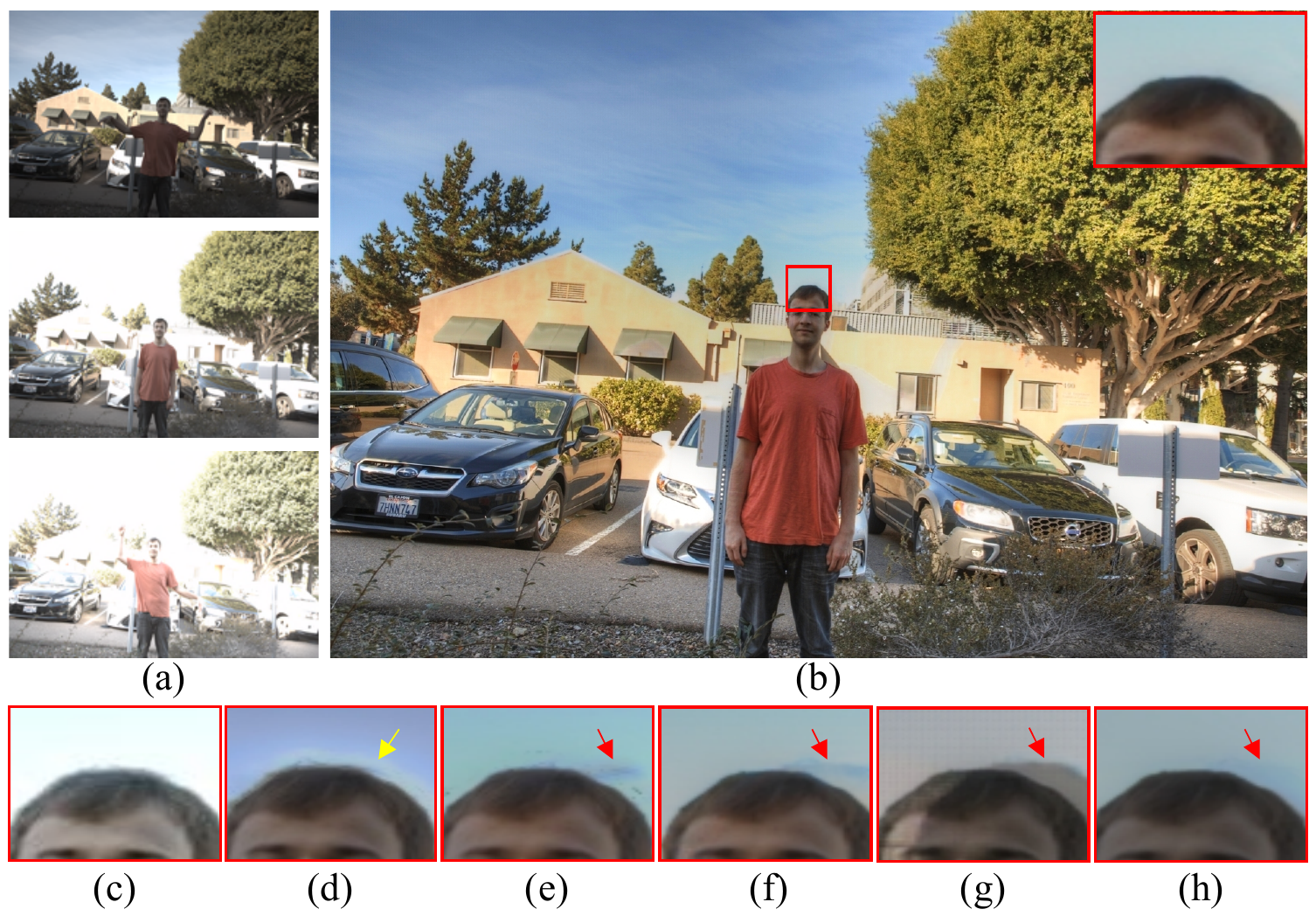}\\
   \caption{LDR images with different exposures are shown in (a), and our result is shown in (b). (c) Result of Hu~\emph{et al.}'s method~\cite{hu2013hdr}. (d) Result of Sen~\emph{et al.}'s method~\cite{sen2012robust}. (e) Result of Kalantari~\emph{et al.}'s method~\cite{kalantari2017deep}. (f) Result of Wu~\emph{et al.}'s method~\cite{wu2018deep}. (g) Result of Yan~\emph{et al.}'s method~\cite{yan2020deep}. (h) Result of Niu~\emph{et al.}'s~\cite{niu2021hdr}. The proposed UPHDR-GAN handles moving objects better and generates results with fewer ghosting artifacts.}
   \label{fig:introduction}
\end{figure}

Since its first introduction in 1990s, HDR imaging techniques evolve quickly, whose applications include saliency detection~\cite{wang2020multi} and video compression~\cite{zhang2015high}. Some HDR imaging methods are first proposed to generate the results through two steps: (1) reconstructing an HDR image; (2) applying the tone mapping algorithms for display~\cite{qiu2007learning}.
These methods are not suitable for handling dynamic scenes because they do not consider the misalignments between different input images. Subsequently, Oh~\emph{et al.} proposed a rank minimization algorithm to detect outliers for HDR generation and align input images~\cite{denton2015deep}. Szpak~\emph{et al.} introduced the Sampson distance to estimate the homography matrix and applied the homography to align input images~\cite{szpak2014sampson}. These methods work well when the inputs are aligned properly. However, completely aligning the multi-exposure images is challenging. The aforementioned methods may produce ghosting or blurring artifacts if the alignment process fails to work. To alleviate the problem, some patch-based methods are proposed to generate fully registered image stacks. Sen~\emph{et al.} considered the HDR reconstruction as an optimization that includes the alignment and reconstruction~\cite{sen2012robust}. Hu~\emph{et al.} built new image stacks using  a variant of PatchMatch to handle saturated regions and avoid the ghosting artifacts~\cite{hu2013hdr}. However, the patch-based methods lack robustness and cannot produce satisfactory results for complicated scenes. 

Inspired by the convolutional neural network (CNN), some learning-based methods are introduced to imitate the fusion process. Kalantari~\emph{et al.}~\cite{kalantari2017deep} and Wu~\emph{et al.}~\cite{wu2018deep} adopted similar network architecture but different in the pre-processing. Kalantari~\emph{et al.}~\cite{kalantari2017deep} applied the flow-based pre-processing to align the inputs, while Wu~\emph{et al.}~\cite{wu2018deep} embedded the alignment process into the network.
Yan~\emph{et al.}~\cite{yan2019attention-guided} and Liu~\emph{et al.}~\cite{liu2021adnet} proposed the attention-guided network to tackle the misalignment and handle the saturation simultaneously. However, due to the unreliability of the image registration, these methods also suffer from unavoidable artifacts. There are also some GAN-based methods that introduce the adversarial loss to improve the unsatisfactory regions by creating realistic information~\cite{niu2021hdr}. Different techniques are introduced to improve the fusion performance. However, the most important problem of deep learning-based fusion methods is that they rely heavily on paired inputs and ground truth. 

To relax the constraint of the dataset, we propose a GAN-based fusion method to optimize the network using unpaired dataset, named UPHDR-GAN.
First, compared to famous single-image enhancement methods~\cite{lore2017llnet,ren2018lecarm,li2020underwater,li2021low} 
and some recent GAN-based image fusion methods~\cite{xu2020mef,yang2020ganfuse,niu2021hdr} that are trained on paired datasets, the proposed method trains unpaired datasets and transfers the multi-exposure LDR domain images to HDR domain images. 
The datasets of common deep learning-based methods require the inputs and the ground truth images. However, obtaining HDR ground truth images is difficult and most existing datasets just include the input images. Some recent datasets~\cite{kalantari2017deep} generate the ground truth images according to the inputs, but their variety of the scenes is so limited. Training the model on unpaired dataset can relax the constrain of paired training and broaden the application of the dataset.
Second, unlike some methods that are designed for unpaired datasets mainly concentrate on processing single-input, our method is a multi-input method with the consideration of moving objects. 
For example, CycleGAN~\cite{zhu2017unpaired} is designed for training unpaired datasets and processing single-input. The CycleGAN is not suitable for fusing multi-exposure inputs
because the forward process (composing multi-exposure images into an HDR output) may be learned properly, while the backward process (decomposing the HDR image into the multi-exposure images) may not converge successfully. The forward process and the backward process in CycleGAN are interactive. Therefore, the forward process will be influenced if the backward process cannot work satisfactorily.
Even considering multi-input, simply concatenating multi-exposure inputs will result in severe ghosting. 

In contrast, the UPHDR-GAN designs specific modules to solve such problems and produce informative HDR outputs with fewer ghosting artifacts. First, we introduce the initialization phase to maintain the content information between the reference and the output. The initialization phase totally avoids ghosting because it just transfers the reference images to HDR domain. Second, we improve the common adversarial loss to generate images with sharp edges (Fig.~\ref{fig:pipeline} (b)). Third, when fusing the information from the under- and over-exposure images, the min-patch training module (Fig.~\ref{fig:pipeline} (c)) is adopted to detect and handle the ghosting artifacts.
The comparison results with several de-ghosting methods are shown in Fig.~\ref{fig:introduction}.
The comparison methods have diverse artifacts, while our UPHDR-GAN handles the dynamic objects properly with the balance of the HDR transformation and content preservation.

In summary, the main contributions include:
\begin{itemize}
\setlength{\itemsep}{0pt}
\setlength{\parsep}{0pt}
\setlength{\parskip}{0pt}
\item  We proposed a GAN-based multi-exposure HDR fusion network, which relaxes the constraint of paired training dataset and learns the mapping between input and target domains. To our best knowledge, this work is the first GAN-based approach for unpaired HDR reconstruction.
\item  The proposed method can not only be trained on unpaired dataset but generate HDR results with fewer ghosting artifacts. We apply the modified GAN loss, the initialization phase and the min-patch training module to avoid ghosting and improve the image quality.
\item We provided comprehensive comparisons with several leading methods. The results demonstrate that the proposed UPHDR-GAN outperforms existing methods and works well on challenging cases.
\end{itemize}

\section{Related Works}

\subsection{HDR Imaging}
HDR imaging has been extensively researched over the past decades. Existing HDR imaging methods can be mainly divided into two groups, static and dynamic scene methods.
\paragraph{Static scene methods} 
Debevec~\emph{et al.} first proposed to fuse different exposure images to an HDR image~\cite{debevec1997recovering}.
The original approaches produced spectacular results for static cameras and static scenes. Some variants are then introduced by generating disparity maps or using neural networks~\cite{Var2008gradient,sun2010hdr}. Sun~\emph{et al.} computed the disparity map first and applied them to compute the camera response function~\cite{sun2010hdr}. 
Hashimoto~\emph{et al.} developed hard-to-view or nonviewable features and content of color images by a new tone reproduction algorithm~\cite{Var2008gradient}. 
There are also numerous static fusion methods that do not generate HDR outputs but directly obtain informative LDR results~\cite{mertens2007exposure,ma2019deep,li2021detail,wang2021unfusion}. 
Li~\emph{et al.} incorporated the edge-preserving factors into the fusion method to preserve the details~\cite{li2021detail}.
Wang~\emph{et al.}~\cite{wang2021unfusion} presented a unified multi-scale densely connected fusion network to fuse the infrared and visible images.
However, due to the lack of an explicit detection for the dynamic objects, the aforementioned methods are unaware of any motion in the scene, so as to be suitable for static scenes only.
\paragraph{Dynamic scene methods}
Many de-ghosting algorithms are introduced to solve the problem that static methods are not applicable for many scenes~\cite{banterle:2017,Tursun2015The}. Some methods compute weight maps of input images and eliminate the moving contents together~\cite{jacobs2008automatic,li2020multi-exposure}. Complementary, some methods merge images first and resolve ghosting of the results~\cite{raman2011reconstruction}. The misaligned pixels often appear in such methods so that they usually fail to fully utilize available content to generate HDR images.
There are also some methods that are applying energy optimization to maintain image consistency or model the noise distribution of color values~\cite{granados2013automatic}.
Besides, some more complicated methods based on optical flow~\cite{li2019hybrid} or patch-based correspondence~\cite{sen2012robust,hu2013hdr} are proposed to achieve more accurate image registration. Li~\emph{et al.} applied the optical flow to roughly align the multi-exposure images which are captured by hand-held cameras and then used the patch-based optimization to obtain full-aligned inputs~\cite{li2019hybrid}.
Sen~\emph{et al.} integrated alignment and reconstruction in a patch-based energy minimization through an HDR image synthesis equation~\cite{sen2012robust}.
Hu~\emph{et al.} built new image stacks using  a variant of PatchMatch to handle saturated regions and avoid the ghosting artifacts~\cite{hu2013hdr}. Although flow-based methods are able to align images with complex motions, they usually suffer from deformations in the regions with no correspondences, due to occlusions caused by parallax or dynamic contents. On the other hand, patch-based methods sometimes produce excellent results, while they are less efficient and usually fail in large motions and saturated regions. To overcome above issues, some deep learning approaches have been developed recently~\cite{kalantari2017deep,wu2018deep,yan2019attention-guided,yan2020deep}. The deep learning methods can obtain information from the training process to compensate for image regions. However, each of these methods only addresses part of the issues and needs paired data to optimize the network. We propose UPHDR-GAN to comprehensively handle existing issues, including solving ghosting artifacts and relaxing the constrain of paired data.

\subsection{GAN-based Fusion}
GAN was proposed by Goodfellow~\emph{et al.}~\cite{goodfellow2014generative}, which has achieved impressive results in image blending~\cite{wu2019gp}, image generation~\cite{lu2018image,yuan2019bridge}, 
image style transfer~\cite{li2020sdp}, and solving jigsaw puzzles~\cite{li2021jigsawgan}.
Generally, the inputs of common GAN-based methods are noise or a single image.
Obtaining information from multi-inputs is also an important research topic~\cite{perera2018in2i:,joo2018generating}.
Guo~\emph{et al.} introduced a GAN-based multi-focus image fusion system, which utilized the generator to produce desired mask maps~\cite{guo2019fusegan:}.
Huang~\emph{et al.} presented an adaptive weight block to determine whether source pixels are focused or not.~\cite{huang2020generative}
Li~\emph{et al.} proposed AttentionFGAN that applies the attention mechanism into the GAN framework and uses the attention features to fuse the infrared and visible image~\cite{li2020attentionfgan}.
Recently, there are some GAN-based methods are proposed to handle multi-exposure images~\cite{xu2020mef,yang2020ganfuse,niu2021hdr}. 
Xu~\emph{et al.} introduced the self-attention mechanism to solve the luminance variety of multi-exposure images~\cite{xu2020mef}. Yang~\emph{et al.} fused the over- and under-exposed image by increasing the number of the discriminators~\cite{yang2020ganfuse}. 
Niu~\emph{et al.} incorporated the adversarial learning and a reference-based residual merging block to solve large motions~\cite{niu2021hdr}. However, these GAN-based methods rely heavily on paired training datasets so that their performances are greatly limited. In comparison, we propose UPHDR-GAN to fuse multi-exposure inputs, which is compatible with unpaired datasets, so that the flexibility and robustness of our proposed network are significantly improved.

\section{Method}

We propose a GAN-based multi-exposure fusion framework, which is the first method designed for handling HDR imaging tasks with unpaired datasets. Like common GAN framework, the generator $G$ transforms inputs of source domain to desired outputs with the characteristics of the target domain, while the discriminator $D$ distinguishes the target domain images from the generated ones to optimize $G$. 
Our collected dataset consists of scenes with and without ground truth. 
By disorganizing the correspondence between the inputs and ground truth, the unpaired training set is obtained.
To better describe the framework, two domain data are collected, including (1) the source LDR domain $X$, which is constituted by a wide diversity of multi-exposure sequences $x=\{x_1, x_2, x_3\}$, and (2) the target domain $Y$, which consists of a collection of HDR images.
We denote their data distributions as $x\sim{p_{{\rm{data}}}}(x)$ and $y\sim{p_{{\rm{data}}}}(y)$, respectively. The proposed UPHDR-GAN can generate HDR images with fewer ghosting artifacts in the absence of paired datasets. 

\subsection{Network Architecture}\label{sec:architecture}

\begin{table}[t]
\centering
\caption{Detailed parameter settings of the network, in which ‘ES’ indicates element-wise sum.}
\label{tab:network}
\setlength{\tabcolsep}{3pt}
\resizebox{1.0\linewidth}{!}{
\begin{tabular}{|c|cc|ccc|c|c|}
\hline
\multicolumn{8}{|c|}{Inputs: 3 $\times$ {[}256, 256, 6{]}}                                                                                                   \\ \hline
\multirow{14}{*}{\textit{G}} & \multicolumn{2}{c|}{\multirow{2}{*}{Module}}                         & \multicolumn{3}{c|}{Conv} & BN      & \multirow{2}{*}{Activation} \\ \cline{4-7} 
                             & \multicolumn{2}{c|}{}                                                & Kernel & Stride & Channel & Channel &            \\ \cline{2-8} 
                             & \multicolumn{1}{c|}{\multirow{5}{*}{Encoder}} & E1                  & 7      & 1      & 64      & 64      & ReLU       \\ \cline{3-8} 
                             & \multicolumn{1}{c|}{}                         & \multirow{2}{*}{E2} & 3      & 2      & 128     & -       & -          \\
                             & \multicolumn{1}{c|}{}                         &                     & 3      & 1      & 128     & 128     & ReLU       \\ \cline{3-8} 
                             & \multicolumn{1}{c|}{}                         & \multirow{2}{*}{E3} & 3      & 2      & 256     & -       & -          \\
                             & \multicolumn{1}{c|}{}                         &                     & 3      & 1      & 256     & 256     & ReLU       \\ \cline{2-8} 
                             & \multicolumn{2}{c|}{\multirow{2}{*}{Residual blocks}}               & 3      & 2      & 256     & 256     & ReLU       \\
                             & \multicolumn{2}{c|}{}                                               & 3      & 1      & 256     & 256     & ES         \\ \cline{2-8} 
                             & \multicolumn{1}{c|}{\multirow{5}{*}{Decoder}} & \multirow{2}{*}{D1} & 3      & 1/2    & 128     & -       & -          \\
                             & \multicolumn{1}{c|}{}                         &                     & 3      & 1      & 128     & 128     & ReLU       \\ \cline{3-8} 
                             & \multicolumn{1}{c|}{}                         & \multirow{2}{*}{D2} & 3      & 1/2    & 64      & -       & -          \\
                             & \multicolumn{1}{c|}{}                         &                     & 3      & 1      & 64      & 64      & ReLU       \\ \cline{3-8} 
                             & \multicolumn{1}{c|}{}                         & D3                  & 7      & 1      & 3       & -       & -          \\ \hline
\multirow{7}{*}{\textit{D}}  & \multicolumn{1}{c|}{\multirow{7}{*}{}}        & C1                  & 3      & 1      & 32      & -       & LReLU      \\ \cline{3-8} 
                             & \multicolumn{1}{c|}{}                         & \multirow{2}{*}{C2} & 3      & 2      & 64      & -       & LReLU      \\
                             & \multicolumn{1}{c|}{}                         &                     & 3      & 1      & 64      & 64      & LReLU      \\ \cline{3-8} 
                             & \multicolumn{1}{c|}{}                         & \multirow{2}{*}{C3} & 3      & 2      & 128     & -       & LReLU      \\
                             & \multicolumn{1}{c|}{}                         &                     & 3      & 1      & 128     & 128     & LReLU      \\ \cline{3-8} 
                             & \multicolumn{1}{c|}{}                         & C4                  & 3      & 1      & 256     & 256     & LReLU      \\ \cline{3-8} 
                             & \multicolumn{1}{c|}{}                         & C5                  & 3      & 1      & 1       & -       & -          \\ \hline
\multicolumn{8}{|c|}{Output HDR ${H_o}$: {[}256, 256, 3{]}}                                                                                                   \\ \hline
\multicolumn{8}{|c|}{Tonemapped HDR $T({H_o})$: {[}256, 256, 3{]}}                                                                                          \\ \hline
\end{tabular}
}
\end{table}

\begin{figure*}[t]
   \centering
   \includegraphics[width=0.98\textwidth]{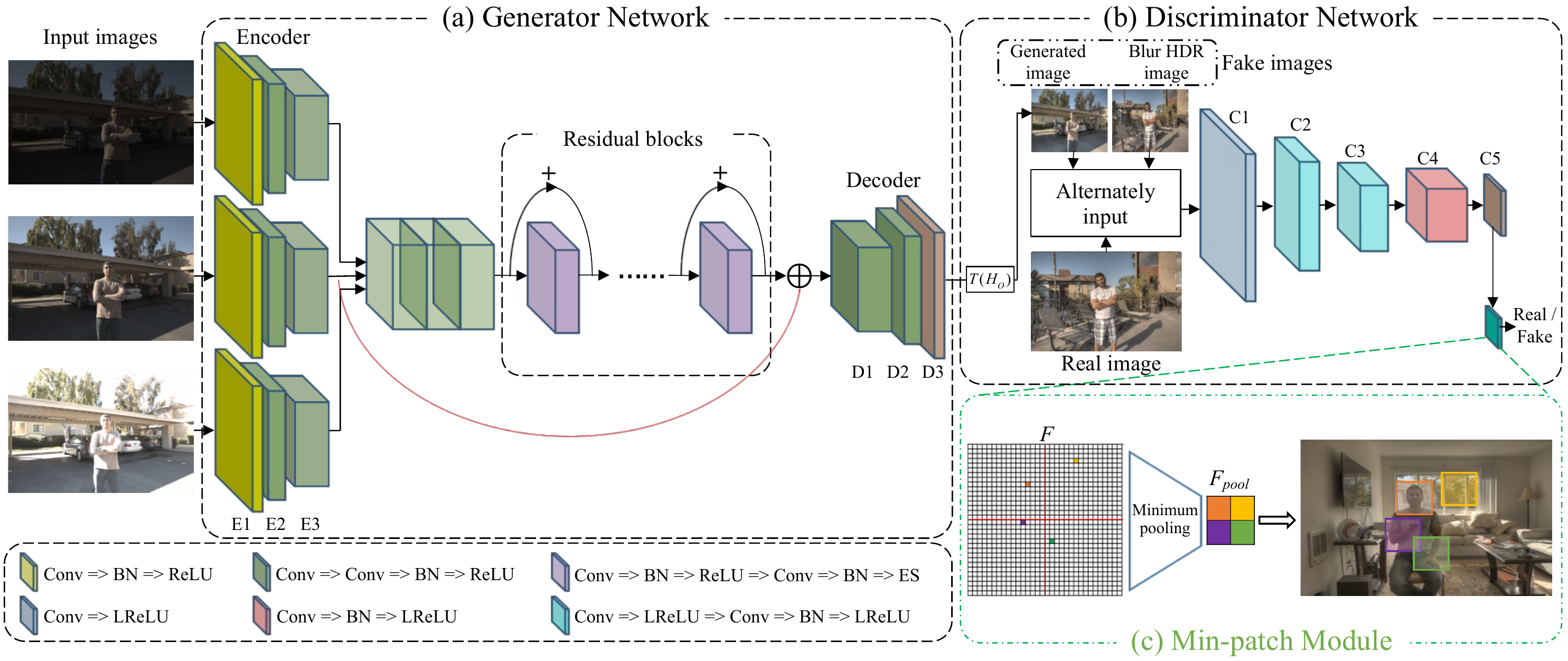}\\
   \caption{
   The proposed method seeks to generate high-quality HDR results with unpaired datasets. 
   \textbf{The generator} first extracts features from multi-exposure inputs using identical down-convolution blocks. The encoder features are then concatenated to be sent to the residual blocks. The decoder recovers the features to informative HDR images through up-convolution blocks.
   \textbf{The discriminator} distinguishes the generated and the real HDR images alternately. \textbf{The min-patch module} concentrates on the strange part of fake images and helps to avoid ghosting artifacts.}
   \label{fig:pipeline}
\end{figure*}

UPHDR-GAN is an images-to-image task with three inputs and one output. The structure of UPHDR-GAN is illustrated in Fig.~\ref{fig:pipeline}.  
The detailed layer configurations of the network architecture are displayed in Table~\ref{tab:network}. 
To improve the efficiency, 
We crop $256 \times 256$ overlapped patches from the training images with a stride of 64 rather than optimizing the model with the full-size images.
The encoder contains three branches and the input size of each branch is $256^2 \times 6$, which is the concatenation of the inputs $x=\{x_1, x_2, x_3\}$ and their mapped HDR images ${H_m} = \{ {H_1},{H_2},{H_3}\}$. ${H_m}$ is obtained using a simple gamma encoding:
\begin{equation}
{H_i} = \frac{{{x_i}^\gamma }}{{{t_i}}},\hspace{.2cm}\gamma  > 1
\end{equation}
where ${x_i}$ is the input image and ${t_i}$ is the corresponding exposure time. The LDR images and the mapped HDR images are complementary, where the former one detects the saturation and misalignments, and the latter one facilitates the convergence of the network across LDR images. 

After getting the HDR output ${H_o}$, we add a $\mu$-law~\cite{kalantari2017deep} post-processing to refine the range of generated HDR images because computing the loss functions on the tone-mapped HDR images is more effective:
\begin{align}\label{eq:u-law}
T({H_o}) = \frac{{\log (1 + \mu {H_o})}}{{\log (1 + \mu )}}
\end{align}
where ${H_o}$ is the output HDR image and real HDR image respectively, $\mu$ represents the amount of compression and is set to 5,000 in our implementation.

\subsubsection{Generator} 
The generator network is composed of the encoder, the residual blocks and the decoder. Specifically, the encoder 
consists of three convolutional blocks: E1, E2 and E3, as described in Table.~\ref{tab:network}. Useful signals are extracted in the encoder process and used for following residual blocks to explore high-level features.
Two transposed convolutional blocks (D1 and D2) and a convolutional layer (D3) constitute the decoder to recover the features to output images.

\subsubsection{Discriminator}
The discriminator is complementary to the generator. PatchGAN~\cite{isola2017image} is applied to classify the image patch
rather than a full image. We crop $70 \times 70$ overlapped patches from generated HDR images and real HDR images to train the patch-based discriminator. 
However, not all regions in the patch contribute to the discriminator optimization during training.
If the generator produces images with regions that are strange and different from the real images,
the special regions can be considered as undesirable ghosting artifacts. Paying more attention to the strangest parts is essential. 

\subsubsection{Min-patch Module}
We introduce the min-patch training module (Fig.~\ref{fig:pipeline} (c)) at the end of the PatchGAN. The implementation of min-patch training is to add an optional minimum pooling layer to the final output of the discriminator~\cite{joo2018generating}.
We define $F$ to represent the features after the `C5' convolutional layer in the discriminator. When training the discriminator, conventional PatchGAN is applied and the network is optimized with $F$. When training the generator, we add the minimum pooling layer after the `C5' convolutional layer. The features after the minimum pooling layer ($F_{pool}$) are used to compute the loss.
The generator is optimized with ${F_{pool}}$, which plays a vital role in detecting the most important parts of the generated images, such as the error parts or strange parts. The discriminator distinguishes the real image from the fake image using common PatchGAN and is trained with $F$. In our implementation, the size of features $F$ after `C5' convolutional layer is $64 \times 64$. 
We use $16 \times 16$ minimum pooling for the min-patch training module and output features $F_{pool}$ with size $4 \times 4$ to optimize the generator.

\subsection{Loss Function}\label{sec:loss}

As GAN is a min-max optimization system, the proposed UPHDR-GAN optimizes the following equation to strike a balance between the generator and the discriminator:
\begin{equation}\label{eq:minmax}
{G^*},{D^*} = \arg \mathop {\min }\limits_{G} \mathop {\max }\limits_D L(G,D)
\end{equation}

Based on HDR imaging properties, the objective function is designed to have the following two items: (1) the GAN loss ${L_{{\rm{GAN}}}}(G,D)$ to achieve desired transformation to convert multi-exposure inputs into HDR outputs; (2) the content loss ${L_{con}}(G)$ to preserve the image semantic information during HDR transformation. The full loss function is:
\begin{equation}\label{eq:loss}
L(G,D) = {L_{{\rm{GAN}}}}(G,D) + {w_{con}}{L_{con}}(G)
\end{equation}
where ${w_{con}}$ is a hyper-parameter to control the relative importance of the content loss, so as to balance the effects of transformation and content preservation.

\begin{figure}[t]
   \centering
   \includegraphics[width=1.0\linewidth]{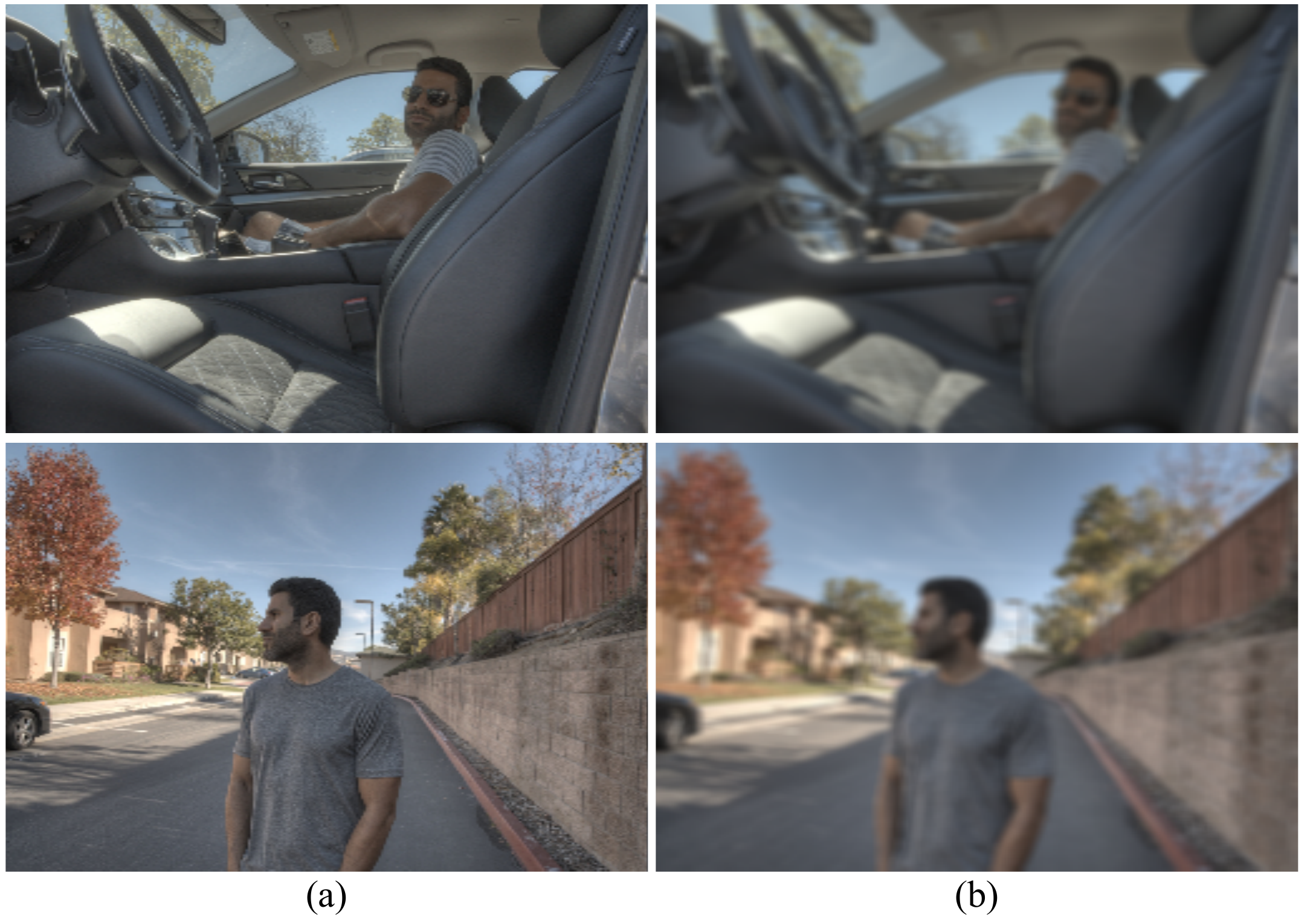}\\
   \caption{Two examples of the blur dataset. (a) The tone-mapped HDR images. (b) Blur results of (a).}
   \label{fig:blur_dataset}
\end{figure}

\subsubsection{GAN Loss}
The GAN loss helps $G$ to generate results similar to the target domain images in the absence of ground truth, and confuses $D$ using the generated HDR images and real HDR images. 
However, applying vanilla GAN loss is insufficient, which cannot preserve the edge and boundary information, while such information is important for HDR images.
For this reason, Chen~\emph{et al.}~\cite{chen2018cartoongan} proposed to confuse $D$ with a blur dataset,
which has been proven useful for the style transformation. The blur dataset is considered as fake images to drive the generator to produce images with clear edges.
Similarly, we also add a blur HDR dataset to facilitate $G$ to generate high-quality output. Specifically, for the target images ${\{ {y_j}\} _{j = 1, ..., M}} \in Y$, we utilize Gaussian filter with kernel size $5 \times 5$ to remove their clear edges and generate the blur dataset ${\{ {b_j}\} _{j = 1, ..., M}} \in B$. We show two examples of the blur dataset in Fig.~\ref{fig:blur_dataset}.
The characteristic of blur edges should be avoided in generated images. Selecting the blur dataset as fake images can help the network produce images without blur edges.
In other words, there are three categories that need to be classified by the discriminator: $G(x)$, $b$ and $y$, among which the generated image $G(x)$ and the blurred HDR image $b$ are fake inputs, and the real HDR image $y$ is real input. 
The modified adversarial loss is designed as:
\begin{equation}\label{eq:gan_loss}
\begin{aligned}
{L_{{\rm{GAN}}}}(G,D)  = & {\mathbb{E}_{y\sim{p_{{\rm{data}}}}(y)}}\big[\log D(y)\big] \\ 
 + & {\mathbb{E}_{x\sim{p_{{\rm{data}}}}(x)}}\big[\log (1 - D(G(x))\big] \\
 + & {\mathbb{E}_{b\sim{p_{{\rm{data}}}}(b)}}\big[\log (1 - D(b))\big] \\ 
\end{aligned}
\end{equation}

We adopt the negative form of the modified adversarial loss in order to use the min-patch training module properly. Conventional adversarial loss minimizes the generator loss while maximizing the discriminator loss. Now, we train the generator to maximize the loss function and the discriminator to minimize the loss function. 
The inverse optimization is specifically designed for the min-patch training module, which is only used when training the generator. The modified 
generator loss tries to maximize the discriminator values after passing the minimum pooling.
The lower discriminator outputs imply the fake patches, which may represent the blur or ghosting regions. The modified generator loss can concentrate on these strange parts by maximizing the lower discriminator values.

\subsubsection{Content Loss}
The GAN loss just ensures the generator produces images that are similar to the real HDR domain images.
The semantic information preservation cannot be guaranteed by using adversarial loss alone. Adding additional constraints for semantic consistency is necessary.
Generally, we select the image with middle-exposure as the reference image, and align images with under- and over-exposure to the reference.
The content loss is defined to constrain the paired middle-exposure input $x_2$ and the generated result $G(x)$ about the semantic similarity.
Instead of using common MSE loss function, the perceptual loss~\cite{johnson2016perceptual} is applied to constrain the content differences, which is formulated as:
\begin{equation}\label{eq:con_loss}
\small
{L_{con}}(G) = {\mathbb{E}_{x\sim{p_{{\rm{data}}}}(x)}} \big[|| VG{G_l}\big(G(x)\big) - VG{G_l}\big(x_2\big)|{|_1}\big]
\end{equation}
where the selection of layers $l$ is important. Larger $l$ will extract high-level features. We utilize the features of the `conv4\_4' layer from the VGG19 network in our method.

The hyper-parameter ${w_{con}}$ is added to balance the adversarial loss and content loss. The adversarial loss works on unpair domain translation, while the content loss constrains pair content preservation.
A larger ${w_{con}}$ destroys the domain transformation and generates results that do not like desired HDR images due to the excessive content preservation from inputs,
while a small ${w_{con}}$ concentrates more on unpaired domain translation and the semantic information of the reference image will be destroyed.
In order to achieve the balance, ${w_{con}}$ is empirically set to be 1.5 at the initial stage. After the training process becomes increasingly stable and the content information from the reference is maintained reasonably, ${w_{con}}$ is gradually decreased to achieve the domain transformation.
${w_{con}}$ is described as:
\begin{equation}\label{eq:con_decay}
w_{con} = w_{con} \times 0.96^{\lfloor N_e/10  \rfloor}
\end{equation}
where $N_e$ is the number of epochs, which is set to 200 in our implementation.

\begin{table*}[t]
\centering
\caption{Detailed source information of our dataset.}
\resizebox{1.0\textwidth}{!}{
\begin{tabular}{
>{\centering\arraybackslash}p{3cm}
>{\centering\arraybackslash}p{12cm}
>{\centering\arraybackslash}p{3cm}}
\toprule
\textbf{Source Name} & \textbf{URL}                                         & \textbf{Number} \\
\midrule
HDReye               & {\texttt{https://mmspg.epfl.ch/downloads/hdr-eye/}}             & 46              \\
Fairchild            & {\texttt{http://rit-mcsl.org/fairchild/HDR.html}}  & 103             \\
EmpaMT               & {\texttt{http://empamedia.ethz.ch/hdrdatabase/index.php}}       & 30              \\
Kalantari~\cite{kalantari2017deep}  & {\texttt{https://cseweb.ucsd.edu/$\sim$viscomp/projects/SIG17HDR}}   & 74 \\
Tursun~\cite{tursun2016objective}   & {\texttt{http://user.ceng.metu.edu.tr/$\sim$akyuz/files/eg2016/index.html}}   & 17 \\
\bottomrule
\end{tabular}
}
\label{tb:hdrsources}
\end{table*}

\begin{figure*}[t]
   \centering
   \includegraphics[width=1.0\linewidth]{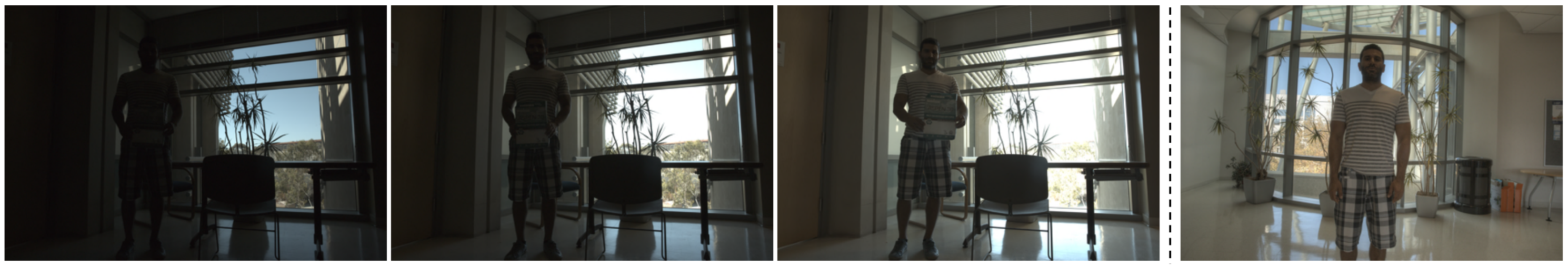}\\
   \caption{An example of the unpaired training dataset, among which three images on the left are the input images, while the rightmost image is the target image in the HDR domain.}
   \label{fig:dataset}
\end{figure*}

\section{Experiments}

The datasets and implementation details are first illustrated in Section~\ref{sec:implementation}.
Comprehensive experiments are then conducted, including quantitative comparisons (Section~\ref{sec:quantitative}), qualitative assessments (Section~\ref{sec:qualitative}) computational complexity (Section~\ref{sec:times}), results on sequences captured by hand-held smartphones (Section~\ref{sec:reallife}) , and ablation studies (Section~\ref{sec:ablation}). 
Specifically, we first compare the proposed method with several methods that can only be applied to fuse static inputs~\cite{mertens2007exposure,li2013image,li2012fast,paul2016multi,ma2019deep,xu2020mef}, and then compare with several classic de-ghosting methods, including two patch-based methods~\cite{sen2012robust,hu2013hdr}, two deep neural network (DNN) mergers with and without optical flow registration, respectively~\cite{kalantari2017deep,wu2018deep}, a non-local network~\cite{yan2020deep}, and a GAN-based method~\cite{niu2021hdr}.
We use the under- and the over-exposed image to produce the results of Xu~\emph{et al.}'s method~\cite{xu2020mef} because their method only takes two inputs.

\subsection{Datasets and Implementation Details}\label{sec:implementation}
The datasets of common deep learning-based multi-exposure fusion methods usually include multi-exposure input images and ground truth HDR image. However, 
obtaining corresponding ground truth HDR images is difficult and most existing datasets just include the input images. 
Moreover, many existing datasets only include static scenes.
Although some of them include moving objects, the dynamic scenes occupy a small proportion.
Kalantari~\emph{et al.} introduced the first HDR dataset, however, the variety of the scenes is so limited~\cite{kalantari2017deep}. Our method relaxes the constraints of paired input and learns the transformation from the source LDR domain to the target HDR domain. The network is trained to fuse multi-exposure inputs in the absence of corresponding ground truth.
We have collected a total of 270 groups of images from various sources, 
as seen in Table~\ref{tb:hdrsources} for the detailed information. 
The ground truth images in the test set are required to compute the quantitative scores. The image sequences from Tursun~\emph{et al.}~\cite{tursun2016objective} and Fairchild do not contain the ground truth images. Therefore, we randomly select dynamic test scenes from other three datasets. Kalatari~\emph{et al.}'s dataset~\cite{kalantari2017deep} only contains dynamic scenes. Twenty static test scenes are randomly selected from remainder two datasets. As for twenty dynamic scenes, 6 sequences originate from the HDReye dataset, 4 sequences originate from the EmpaMT dataset and 10 sequences originate from the Kalantari~\emph{et al.}'s dataset~\cite{kalantari2017deep}. As for twenty static scenes, 11 sequences originate from the HDReye dataset and 9 sequences originate from the EmpaMT dataset. Finally, 40 groups of images are selected as the test set and 230 groups of images are selected as the training set. The test set and the training set are completely distinct. Some of the sequences include approximately 10 multi-exposure inputs, from which we select 3 images with minimum, medium and maximum exposure as training inputs.

By disorganizing the correspondence between the inputs and ground truth, the unpaired training set is obtained. An example of the unpaired training dataset is shown in Fig.~\ref{fig:dataset}.
The training images are first aligned using a homography before they are sent to the network, which is more effective and helps the network concentrates more on the moving objects.
All training images are resized to $1000 \times 1500$.
Then, we crop $256 \times 256$ overlapped patches from the training images with a stride of 64 to improve the training efficiency. The pre-processing will create 54,240 patches.
After that, we utilize the data augmentation, including the flipping and rotation to enrich the training data by 8 times. Finally, the training set consists of 433,920 training patches, which is large enough to encompass all the possibilities and train our architecture.

\begin{figure}[t]
   \centering
   \includegraphics[width=1.0\linewidth]{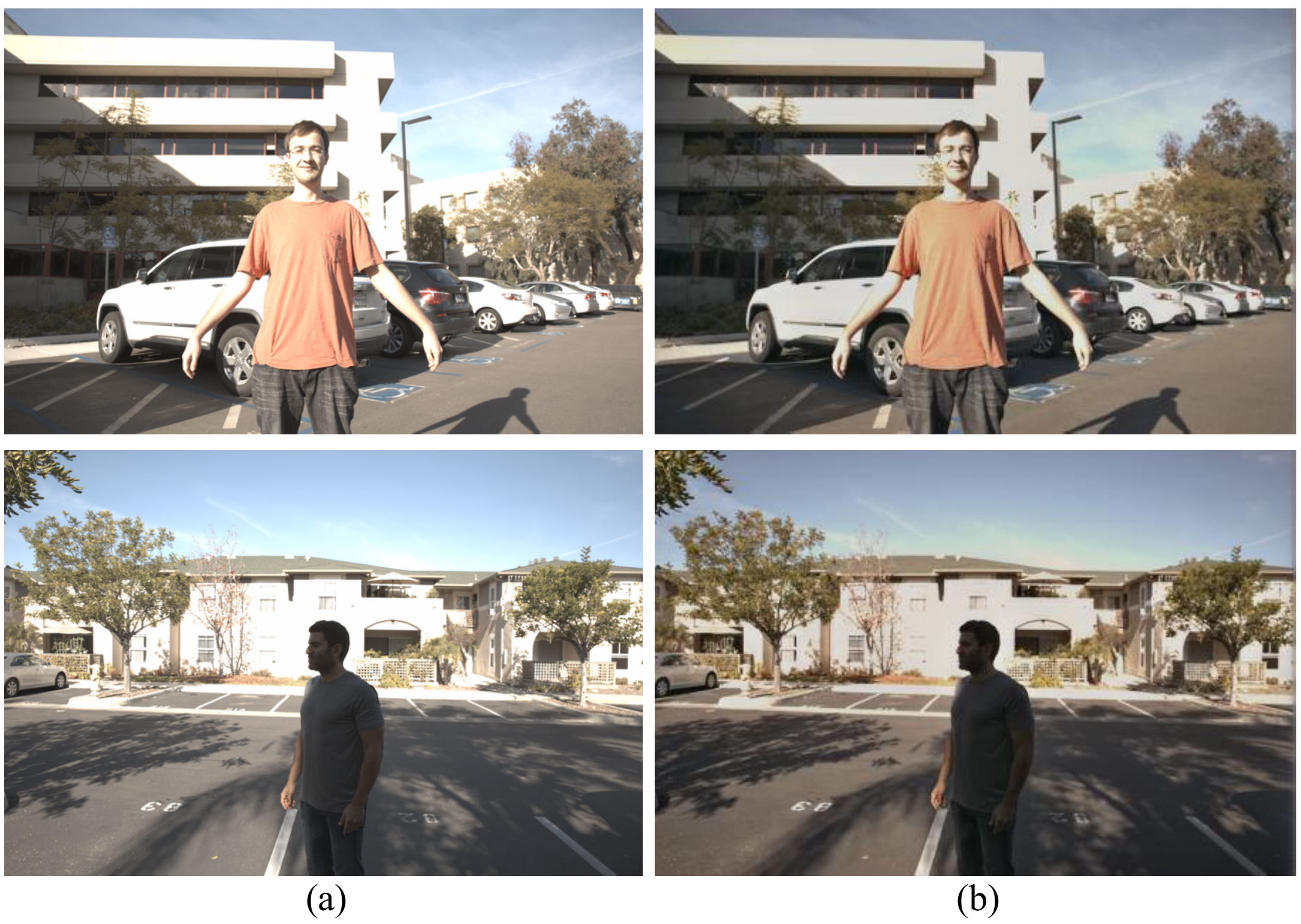}\\
   \caption{Results of the initialization phase. (a) The middle-exposure inputs. (b) The generated results after the pre-training with 10 epochs.}
   \label{fig:initialization}
\end{figure}

\begin{table*}[t]
\centering
\caption{Quantitative comparison of UPHDR-GAN with the comparison methods on twenty static scenes. The left part shows the comparison results with methods that are suitable for static scenes, and the right part represents the comparison results with methods that are both suitable for static and dynamic scenes.
{\color{red}Red} color indicates the best performance and {\color{blue}blue} color indicates the second-best results. The best results of static methods are underlined.}
\label{tab:static_quantitative}
\resizebox{0.99\textwidth}{!}{
\begin{tabular}{c|cccccc|cccccc|c}
\toprule
Methods & \begin{tabular}[c]{@{}c@{}}Mertens\\\cite{mertens2007exposure}\end{tabular} & \begin{tabular}[c]{@{}c@{}}Li2012\\\cite{li2012fast}\end{tabular} & \begin{tabular}[c]{@{}c@{}}Li2013\\\cite{li2013image} \end{tabular} & \begin{tabular}[c]{@{}c@{}}Paul2016\\\cite{paul2016multi} \end{tabular}  & \begin{tabular}[c]{@{}c@{}}Ma2019\\\cite{ma2019deep}\end{tabular} & \begin{tabular}[c]{@{}c@{}}Xu2020\\\cite{xu2020mef} \end{tabular} & \begin{tabular}[c]{@{}c@{}}Sen\\\cite{sen2012robust} \end{tabular}  & \begin{tabular}[c]{@{}c@{}}Hu\\\cite{hu2013hdr} \end{tabular} & \begin{tabular}[c]{@{}c@{}}Kalantari\\\cite{kalantari2017deep} \end{tabular} & \begin{tabular}[c]{@{}c@{}}Wu\\\cite{wu2018deep} \end{tabular} & \begin{tabular}[c]{@{}c@{}}Yan\\\cite{yan2020deep} \end{tabular} & \begin{tabular}[c]{@{}c@{}}Niu\\\cite{niu2021hdr} \end{tabular} & Ours   \\
\midrule
PSNR $\uparrow$            & 29.817  & 30.289 & 31.020 & 31.876   & \underline{33.469} & 32.582 & 39.105    & 32.192    & 40.008          & 39.505    & 39.994    & {\color{red}40.637}    & {\color{blue}40.601} \\
SSIM $\uparrow$            & 0.9511  & 0.9527 & 0.9575 & 0.9584   & \underline{0.9675} & 0.9651 & 0.9664   & 0.9655   & 0.9701           & 0.9681   & 0.9692    &  {\color{blue}0.9715}  &  {\color{red}0.9717} \\
HDR-VDP-2.2 $\uparrow$     & 54.904  & 53.432 & 56.058 & 56.935   & \underline{57.910} & 54.151 & 56.345     & 55.973    & 59.782          & 60.155    & 58.362     & {\color{blue}61.348}    & {\color{red}61.916} \\
TMQI $\uparrow$            & 0.854   & 0.859  & 0.871  & 0.874    & \underline{0.887}  & 0.872  & 0.882   & 0.878    & 0.889          &  0.890   & 0.887    & {\color{blue}0.893}    & {\color{red}0.895}  \\ 
\bottomrule
\end{tabular}
}
\end{table*}

We implement UPHDR-GAN in PyTorch and the model is trained on an NVIDIA RTX 2080Ti GPU for 200 epochs. The entire training process costs 2 days on average. Adam optimizer is selected to iterate the network. The learning rate of the generator and the discriminator is set to $2.0{\rm{ \times 1}}{{\rm{0}}^{{\rm{ - 4}}}}$ and $1.0{\rm{ \times 1}}{{\rm{0}}^{{\rm{ - 4}}}}$, respectively. 
We introduce an initialization phase to help the convergence and guide the network to learn the correct domain transformation.
In initialization, the generator is designed to reconstruct the semantic information of middle-exposure input and ignore the domain translation. For this purpose, the generator $G$ is pre-trained using merely the content loss ${L_{con}}$. Two examples are presented in Fig.~\ref{fig:initialization} that include the input images and the results after pre-training. 
Ablation experiments of the initialization phase are also performed in Section~\ref{sec:ablation}.
The initialization phase contributes to controlling the over-exposed regions and enriching the overall colors. Moreover, the network properly reconstructs the content information of middle-exposure input. Since we select the  middle-exposure image as the reference, the initialization also helps to avoid ghosting.

\subsection{Quantitative Comparisons}\label{sec:quantitative}

Although the proposed UPHDR-GAN can efficiently fuse multi-exposure images without ground truth, we select the test set for quantitative comparisons from paired datasets that include multi-exposure inputs and HDR images. As the ground truth is available, we can conduct various quantitative evaluations and comparisons.
As for the comparisons with static scenes, we compute four metrics, including the PSNR values~\cite{hore2010image}, the SSIM values~\cite{wang2004image}, the HDR-VDP-2.2 scores~\cite{narwaria2015hdr} and the tone mapped image quality index  (TMQI) scores~\cite{yeganeh2012objective}.
The PSNR value approaches infinity as the MSE approaches zero and a higher PSNR value provides a higher image quality. 
The SSIM is considered to be correlated with the quality perception of the human visual system~\cite{wang2004image}. HDR-VDP-2.2 is a calibrated objective method that can tackle both HDR and LDR signals~\cite{narwaria2015hdr}. The TMQI score combines the multi-scale signal fidelity measure and a naturalness measure to evaluate the tome mapped images~\cite{yeganeh2012objective}.
As for the comparisons with dynamic scenes, we further compute the PU-PSNR and PU-SSIM values~\cite{mantiuk2021pu21} with 1,000 $cd/m^{2}$ display, which represents current commercial HDR display technology.
The two perceptually uniform (PU)-encoding metrics convert absolute HDR linear color values into approximately perceptually uniform values and expect that the values in images correspond to the luminance emitted from the HDR display.
The higher PSNR, SSIM, HDR-VDP-2.2, TMQI, PU-PSNR and PU-SSIM scores indicate better image quality.
The quantitative comparison results are presented in Table~\ref{tab:static_quantitative} and~\ref{tab:dynamic_quantitative}.

Twenty static scenes and twenty dynamic scenes, which include multi-inputs and corresponding ground truth, are collected as the test set for quantitative comparisons. The test set is completely distinct from the training set to ensure the evaluation is fair.
The proposed method is first compared with several classic methods that can only be applied to fuse static inputs~\cite{mertens2007exposure,li2012fast,li2013image,paul2016multi,ma2019deep,xu2020mef}. The left part in Table~\ref{tab:static_quantitative} displays the quantitative comparison results with the static methods. 
Some of static methods fuse multi-exposure inputs with the absence of ground truth, and therefore resulting in lower scores when computing the evaluation metrics between the generated image and the ground truth. 
The comparison results with several de-ghosting methods~\cite{sen2012robust,hu2013hdr,kalantari2017deep,wu2018deep,yan2020deep,niu2021hdr} on these static scenes are then reported in the right part of Table~\ref{tab:static_quantitative}. These methods are designed for handling sequences with moving objects, which can solve the slight movements (such as the moving leaves caused by the wind and the flowing water) and obtain higher scores than the aforementioned static methods.
The proposed UPHDR-GAN abandons the constraint of ground truth, but can extract information from the target HDR dataset, hence providing results with better PSNR, SSIM, HDR-VDP-2.2 and TMQI values on average.

\begin{table*}[t]
\centering
\caption{Quantitative comparison of UPHDR-GAN with the dynamic methods on twenty dynamic scenes.
{\color{red}Red} color indicates the best performance and {\color{blue}blue} color indicates the second best results.}
\label{tab:dynamic_quantitative}
\resizebox{0.59\textwidth}{!}{
\begin{tabular}{cccccc}
\toprule
Methods                                  & PSNR $\uparrow$    & SSIM $\uparrow$ & PU-PSNR $\uparrow$    & PU-SSIM $\uparrow$       & HDR-VDP-2.2 $\uparrow$  \\
\midrule
Sen~\cite{sen2012robust}             & 40.924   & 0.9806  & 41.856 & 0.9832    & 57.249 \\
Hu~\cite{hu2013hdr}                  & 34.785   & 0.9725   & 38.604 & 0.9760    & 56.427 \\
Kalantari~\cite{kalantari2017deep}   & 42.532   & 0.9871  & 40.710 & 0.9821    & 61.988 \\
Wu~\cite{wu2018deep}                 & 41.660   & 0.9844   & 41.054 & 0.9854   & 62.345 \\
Yan~\cite{yan2020deep}   & 42.321  & 0.9869    & 40.942 & {\color{blue}0.9855}  & 59.417  \\
Niu~\cite{niu2021hdr}   & {\color{red}43.113}  & {\color{blue}0.9877}    & {\color{blue}41.969}  & {\color{red}0.9860}  & {\color{blue}63.050}         \\
\midrule
Ours                                 & {\color{blue}43.005}   & {\color{red}0.9880} & {\color{red}42.115} & {\color{red}0.9860}   &  {\color{red}63.542} \\
\bottomrule
\end{tabular}
}
\end{table*}

\begin{figure*}[!ht]
   \centering
   \includegraphics[width=1.0\textwidth]{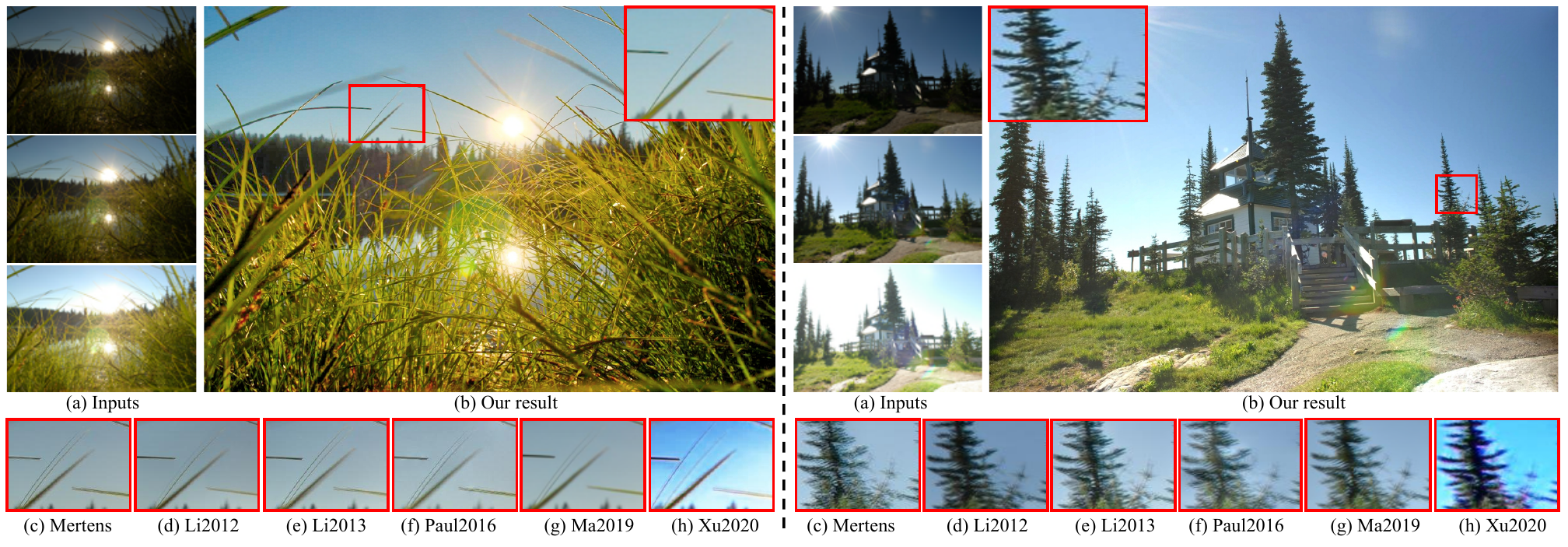}\\
   \caption{Visual comparisons with several representative static methods. (a) Input images. (b) Our result. (c) Result of Mertens~\emph{et al.}'s method~\cite{mertens2007exposure}. (d) Result of Li~\emph{et al.}'s method~\cite{li2012fast}. (e) Result of Li~\emph{et al.}'s method~\cite{li2013image}. (f) Result of Paul~\emph{et al.}'s method~\cite{paul2016multi}. (g) Result of Ma~\emph{et al.}'s method~\cite{ma2019deep}. (h) Result of Xu~\emph{et al.}'s method~\cite{xu2020mef}.}
   \label{fig:comp_static}
\end{figure*}

\begin{figure*}[!ht]
   \centering
   \includegraphics[width=1.0\textwidth]{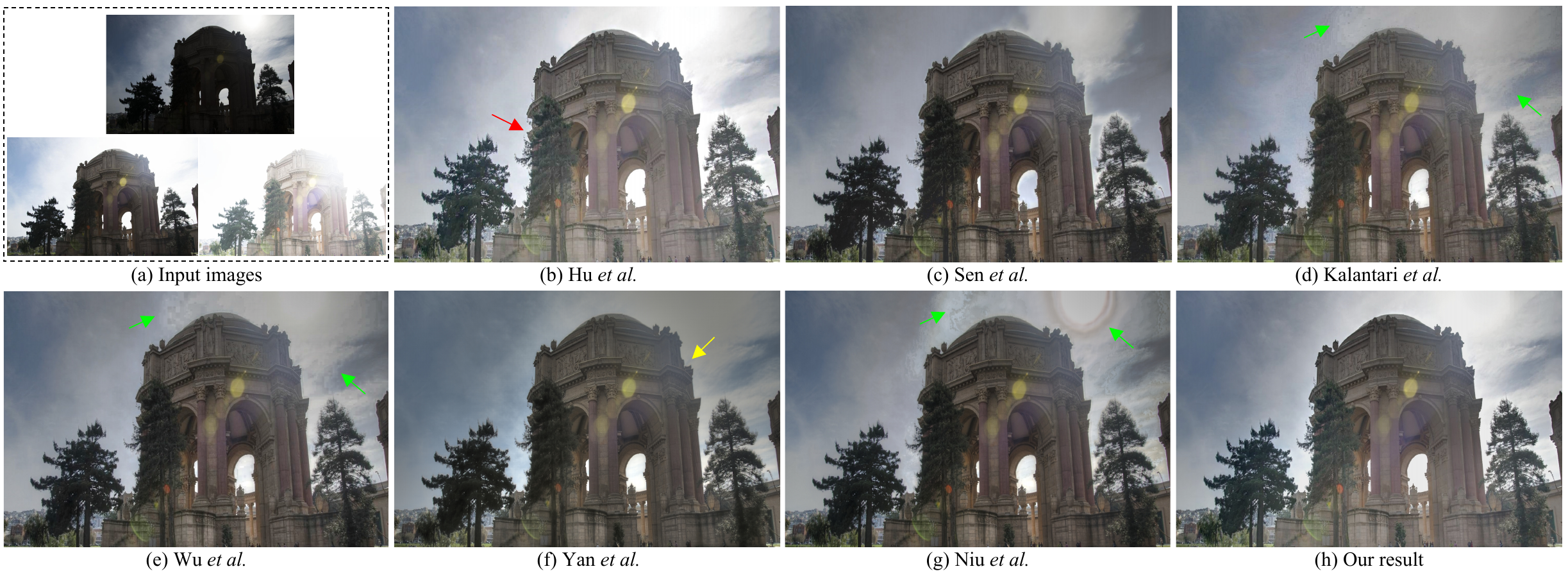}\\
   \caption{Visual comparisons with de-ghosting methods. (a) Input images. (b) Result of Hu~\emph{et al.}'s method~\cite{hu2013hdr}. (c) Result of Sen~\emph{et al.}'s method~\cite{sen2012robust}. (d) Result of Kalantari~\emph{et al.}'s method~\cite{kalantari2017deep}. (e) Result of Wu~\emph{et al.}'s method~\cite{wu2018deep}. (f) Result of Yan~\emph{et al.}'s method~\cite{yan2020deep}. (g) Result of Niu~\emph{et al.}'s~\cite{niu2021hdr}. (h) Our result. Note that, Hu~\emph{et al.}'s method~\cite{hu2013hdr} produces noise around the building in (b). Please zoom in for details.
   }
   \label{fig:comp1}
\end{figure*}

Table~\ref{tab:dynamic_quantitative} exhibits the comparison results of UPHDR-GAN with several de-ghosting methods~\cite{sen2012robust,hu2013hdr,kalantari2017deep,wu2018deep,yan2020deep,niu2021hdr} on twenty dynamic scenes. Two patch-based methods~\cite{sen2012robust,hu2013hdr} generate the registered image stacks according to the patch match-oriented optimization.
Kalantari~\emph{et al.}~\cite{kalantari2017deep} and Wu~\emph{et al.}~\cite{wu2018deep} obtain HDR results through deep neural networks. 
Yan~\emph{et al.} use the non-local correlation to tackle the ghosting artifacts~\cite{yan2020deep}. 
Niu~\emph{et al.} introduce the adversarial loss to improve the unsatisfactory regions by creating realistic information~\cite{niu2021hdr}.
These deep learning-based algorithms have demonstrated significant performance advantages over patch-based methods. However, the deep learning-based methods are not sensitive to large motions and lack robustness.
These comparison methods focus on fusing the multi-exposure images but cannot handle the dynamic objects well, which affects their performance. On the contrary, the proposed initialization phase totally avoids ghosting because it just transfers the reference images to the HDR domain. Then, when fusing the information from the under- and over-exposure images, the min-patch training module helps to detect and avoid ghosting artifacts. Overall,
by incorporating the initialization phase and the min-patch training module, our method owns superior performance.

\subsection{Qualitative Comparisons}\label{sec:qualitative}
In this section, our method is first compared with~\cite{mertens2007exposure,li2013image,li2012fast,paul2016multi,ma2019deep,xu2020mef} on static scenes (Fig.~\ref{fig:comp_static}). The comparison methods are mature enough to handle images that are static, but ignore the tiny motions, such as the moving leaves caused by wind. The comparison methods produce the ghosting artifacts in the left case in Fig.~\ref{fig:comp_static}, which are caused by the slight movements of the leaves. 
Some of them design specific strategies to detect and solve the dynamic contents, such as guided filtering~\cite{li2013image}. However, the results are still unsatisfactory.
In the right case, the static methods suffer from the blurring artifacts around the tree.
Xu~\emph{et al.} solely obtained information from the under- and over-exposed images~\cite{xu2020mef}, which leads to the mediocre result with color deviation.

\begin{figure*}[t]
   \centering
   \includegraphics[width=1.0\textwidth]{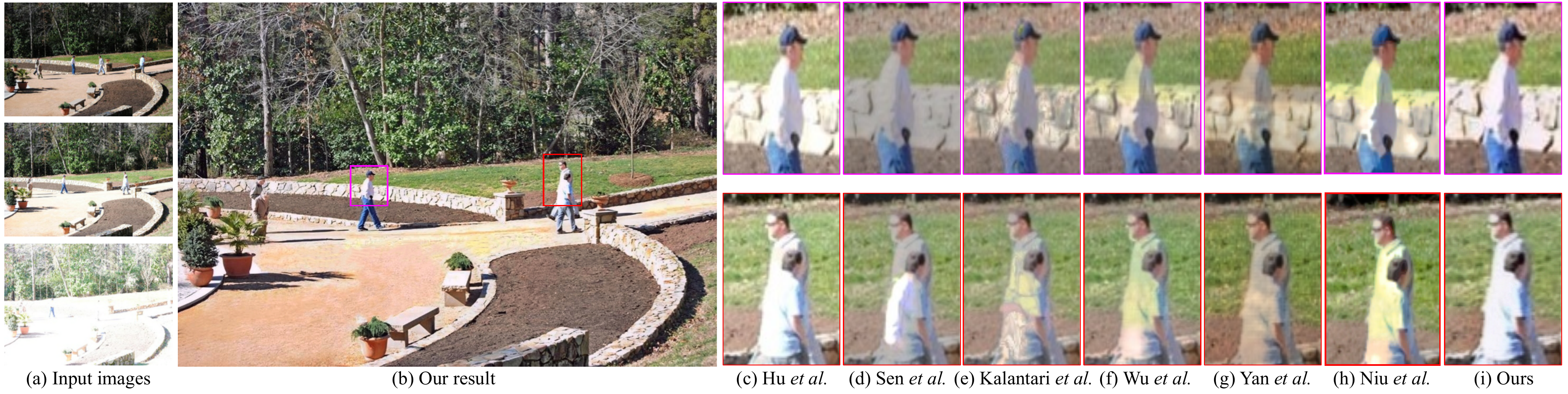}\\
   \caption{Visual comparisons with de-ghosting methods. (a) Input images. (b) Our result. (c) Result of Hu~\emph{et al.}'s method~\cite{hu2013hdr}. (d) Result of Sen~\emph{et al.}'s method~\cite{sen2012robust}. (e) Result of Kalantari~\emph{et al.}'s method~\cite{kalantari2017deep}. (f) Result of Wu~\emph{et al.}'s method~\cite{wu2018deep}. (g) Result of Yan~\emph{et al.}'s method~\cite{yan2020deep}. (h) Result of Niu~\emph{et al.}'s~\cite{niu2021hdr}. (i) Zoomed-in areas of our result. The scene is challenging because there are large foreground motions between input LDR images. The proposed UPHDR-GAN can properly deal with the motions caused by moving people.}
   \label{fig:comp3}
\end{figure*}

\begin{table*}[!ht]
\centering
\caption{The inference time and parameters of different methods on the testing set with size $1000 \times 1500$. The `-' denotes that the patch match-based methods do not have parameters.}
\label{tab:complexity}
\resizebox{0.7\textwidth}{!}{
\begin{tabular}{cccccccc}
\toprule
Methods     & Sen~\cite{sen2012robust}   & Hu~\cite{hu2013hdr}    & Kalantari~\cite{kalantari2017deep} & Wu~\cite{wu2018deep}   & Yan~\cite{yan2020deep}  & Niu~\cite{niu2021hdr}  & PBR-GAN \\ 
\midrule
Environment & CPU   & CPU   & CPU+GPU   & GPU  & GPU  & GPU  & GPU     \\
Time ($s$)        & 61.81 & 79.77 & 29.14     & 0.24 & 0.31 & 0.29 & 0.25    \\
Parameters (M)  & -     & -     & 0.3       & 20.4 & 38.1 & 2.56 & 2.21    \\ 
\bottomrule
\end{tabular}
}
\end{table*}

\begin{figure}[!ht]
   \centering
   \includegraphics[width=1.0\linewidth]{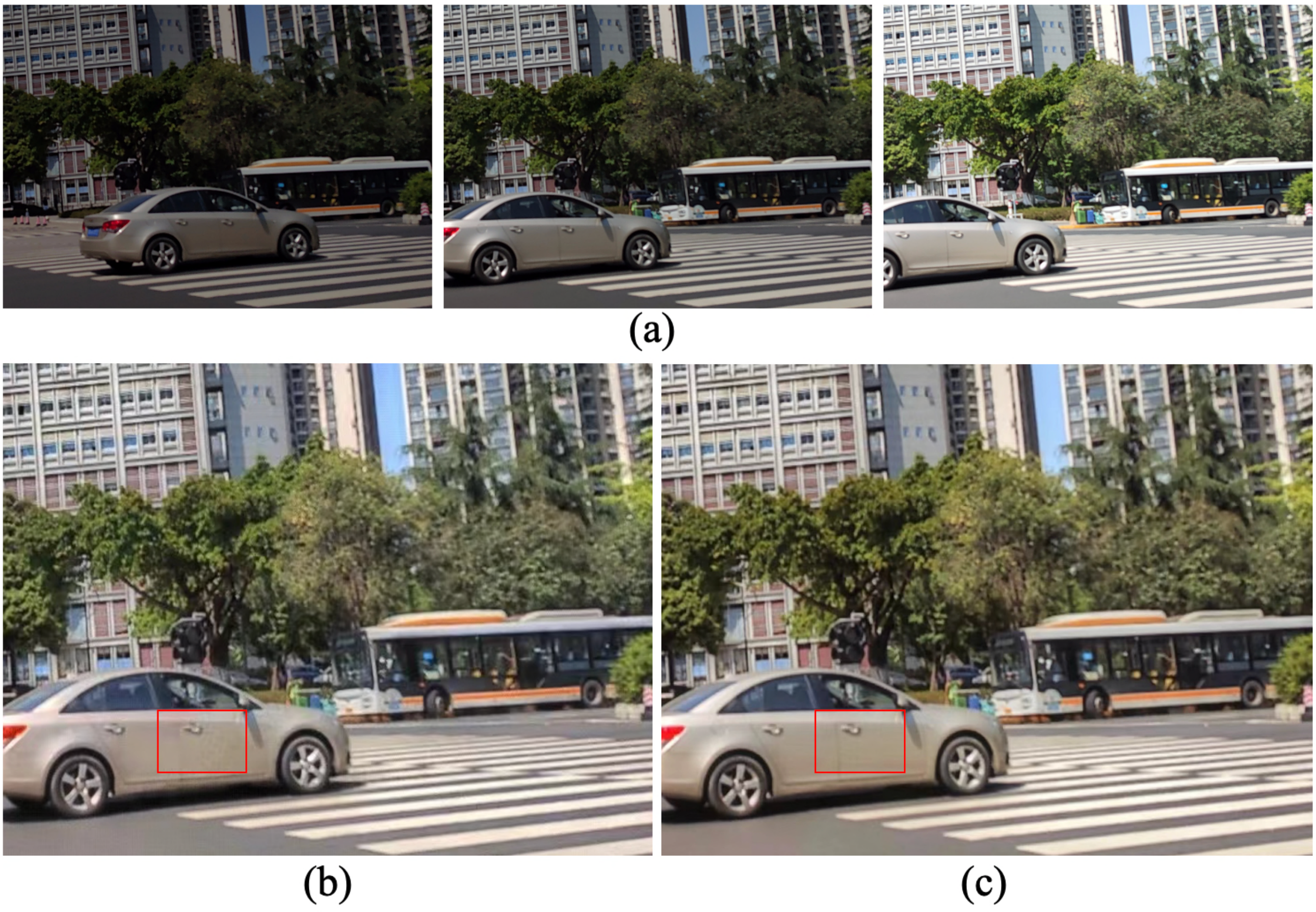}\\
   \caption{Comparisons with Niu~\emph{et al.}'s work~\cite{niu2021hdr} on scene with large motions. (a) Input images with a moving car, which causes large motions. (b) Result of Niu~\emph{et al.}'s work~\cite{niu2021hdr}. (c) Our result.}
   \label{fig:comp_hdrgan}
\end{figure}

\begin{figure}[!ht]
   \centering
   \includegraphics[width=1.0\linewidth]{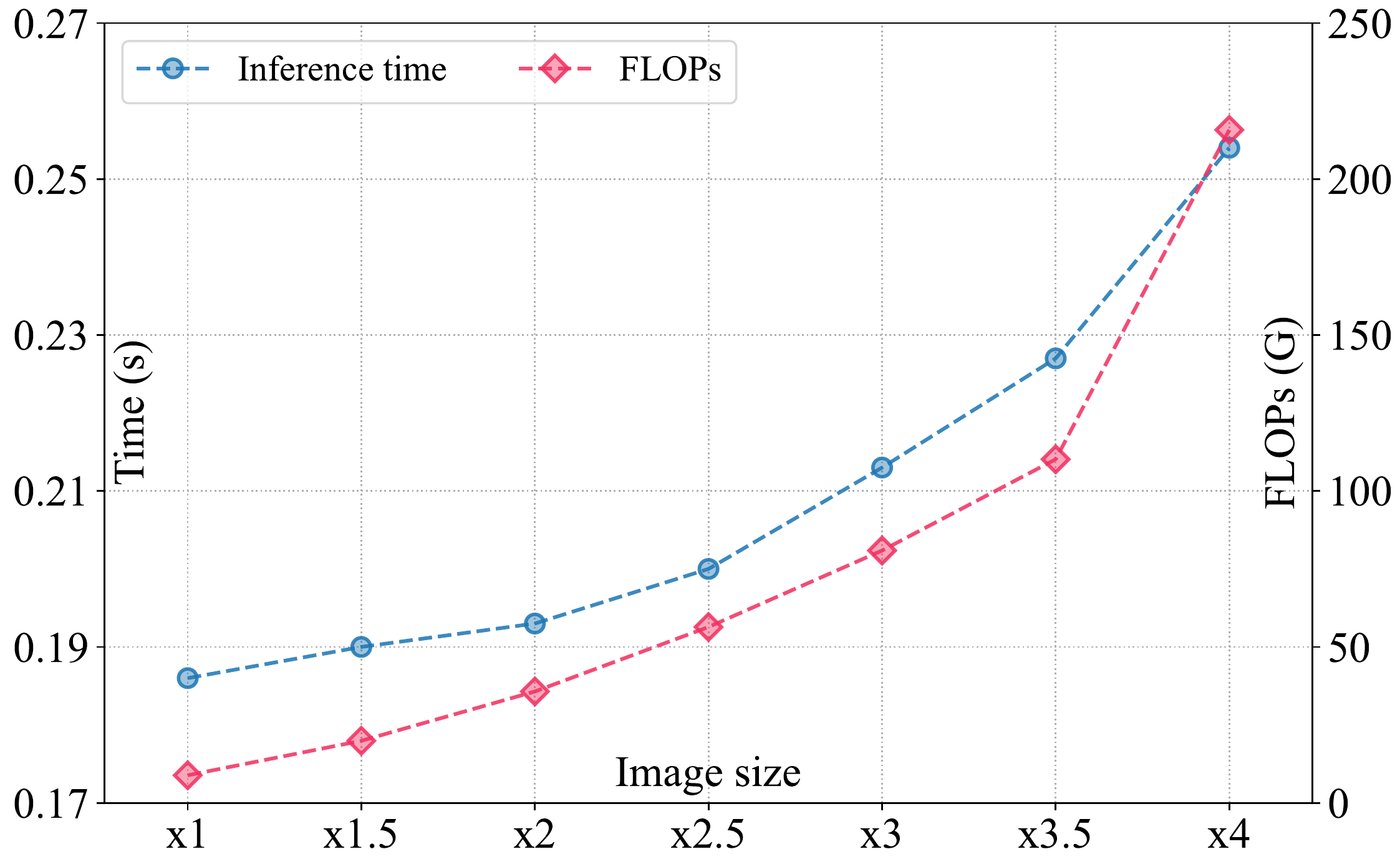}\\
   \caption{The variation trend of the FLOPs and the inference times when selecting test images with different resolutions. The smallest image size in the figure is $256 \times 384$, which is labeled as $\times1$. The image size of $\times1.5$ in the figure is $384 \times 576$. Therefore, the largest image size $\times4$ is $1024 \times 1536$.}
   \label{fig:flops_times}
\end{figure}

\begin{figure*}[t]
   \centering
   \includegraphics[width=1.0\textwidth]{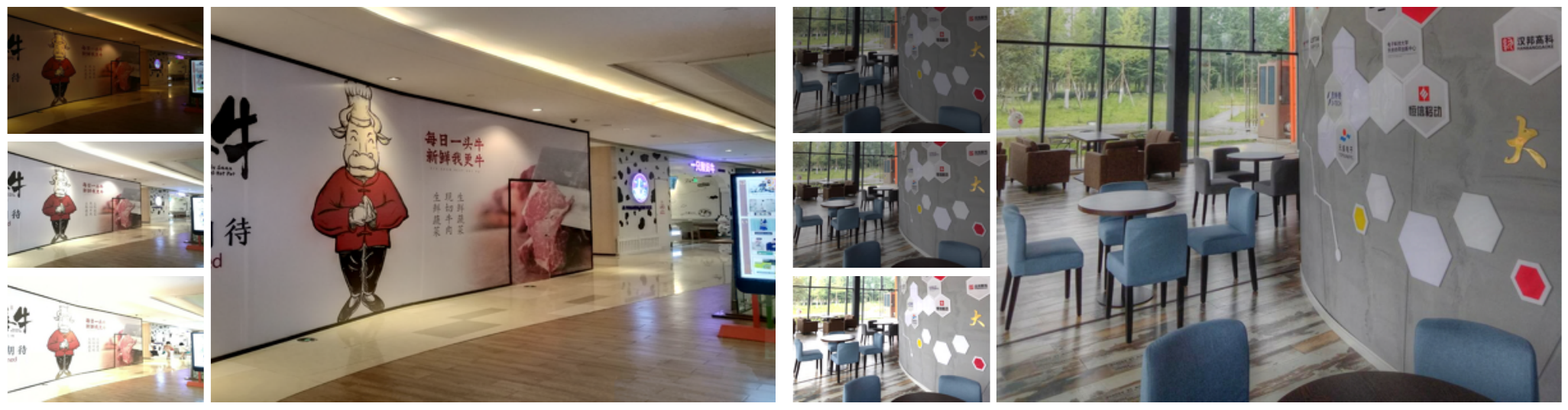}\\
   \caption{The results on real-life scenes captured by HUAWEI Mate 10 smartphones. Each case includes three inputs with different exposures and corresponding results generated by UPHDR-GAN.}
   \label{fig:real_image}
\end{figure*}

Fig.~\ref{fig:comp1} and Fig.~\ref{fig:comp3} show the qualitative comparisons against several state-of-the-art de-ghosting methods~\cite{hu2013hdr,sen2012robust,kalantari2017deep,wu2018deep,yan2020deep,niu2021hdr}.
Two patch-based methods~\cite{hu2013hdr,sen2012robust} tend to generate fully registered input image stacks, but cannot reconstruct the regions with rich textures or large motions.
Hu~\emph{et al.}'s method~\cite{hu2013hdr} generates results with noise around the building (red arrow) in Fig.~\ref{fig:comp1} and unclear edges in Fig.~\ref{fig:comp3}. Sen~\emph{et al.}'s method~\cite{sen2012robust} produces results with serious halo artifacts in Fig.~\ref{fig:comp1} and ghosting artifacts in Fig.~\ref{fig:comp3}.
The deep learning-based methods can obtain information from the training process to compensate for image regions.
However, they only perform well in one way or another. 
Kalantari~\emph{et al.}~\cite{kalantari2017deep} and Wu~\emph{et al.}~\cite{wu2018deep} adopt similar network architecture but different in pre-processing. Kalantari~\emph{et al.}~\cite{kalantari2017deep} apply flow-based pre-processing to align the inputs, while Wu~\emph{et al.}~\cite{wu2018deep} process the alignment and the fusion together. The two methods suffer from similar artifacts, including the problematic transformation in the junction regions of the sky and the cloud (green arrows) in Fig.~\ref{fig:comp1}, and the ghosting artifacts in Fig.~\ref{fig:comp3}. Yan~\emph{et al.} decrease the ghosting artifacts by using the non-local module, which is designed based on the pixel correspondence~\cite{yan2020deep}.
However, their method cannot generate sharp edges (yellow arrow) in Fig.~\ref{fig:comp1} and cannot avoid the ghosting artifacts in Fig.~\ref{fig:comp3}.
Niu~\emph{et al.} incorporated the adversarial learning to produce faithful information in the regions with missing content~\cite{niu2021hdr}.
Their method also suffers from the problematic transformation in the junction regions (green arrows) in Fig.~\ref{fig:comp1} and the unreasonable color reconstruction in Fig.~\ref{fig:comp3}. 
Our method is more sensitive to ghosting artifacts and handles them properly.
We further show the comparisons with Niu~\emph{et al.}'s method on scene with large motions. Fig.~\ref{fig:comp_hdrgan} show the input images (Fig.~\ref{fig:comp_hdrgan} (a)) and results of Niu~\emph{et al.}'s method (Fig.~\ref{fig:comp_hdrgan} (b)) and our method (Fig.~\ref{fig:comp_hdrgan} (c)). Overall, our method achieves comparable result with Niu~\emph{et al.}'s method. Specifically, our method preserves more details than Niu~\emph{et al.}'s method, such as the crevice between two car doors (red box).

\begin{table*}[t]
\centering
\caption{Ablation experiments of different components. {\color{red}Red} color indicates the best performance and {\color{blue}blue} color indicates the second best results.}
\label{tab:ablation}
\resizebox{0.99\textwidth}{!}{
\begin{tabular}{cccccccccc}
\toprule
 & $w_{con}=0.25$   & $w_{con}=0.5$      & $w_{con}=1$ & MSE & w/o. initialization & w/o. min-patch & w/o. blur dataset & only Kalantari's dataset~\cite{kalantari2017deep} & Ours  \\
\midrule
PSNR $\uparrow$ & 36.587 & 39.745 & 41.623 & 36.209 & 41.231 & 40.724 & 42.441 & {\color{blue}42.995} & {\color{red}43.005}\\
SSIM $\uparrow$ & 0.9789 & 0.9803 & 0.9840 & 0.9751 & 0.9837 & 0.9811 & 0.9861 & {\color{red}0.9881} & {\color{blue}0.9880}\\
PU-PSNR $\uparrow$ & 35.512 & 37.559 & 40.023 & 34.725 & 39.574 & 38.827 & 41.721 & {\color{blue}42.107} & {\color{red}42.115}\\
PU-SSIM $\uparrow$ & 0.9766 & 0.9779 & 0.9829 & 0.9758 & 0.9820 & 0.9782 & 0.9852 & {\color{blue}0.9859} & {\color{red}0.9860}\\
HDR-VDP-2.2 $\uparrow$ & 56.842 & 58.196 & 60.018 & 56.131 & 59.877 & 58.624 & 61.596 & {\color{blue}63.491} & {\color{red}63.542}\\
\bottomrule
\end{tabular}
}
\end{table*}

\subsection{Computational Complexity}\label{sec:times}
Computing efficiency is also an important factor for evaluating the fusion performance. The comparisons of inference time and parameters are then conducted. The results for fusion images with size $1000 \times 1500$ on the test set are reported in Table~\ref{tab:complexity}. There is a large difference between different methods. Two patch match-based methods~\cite{sen2012robust,hu2013hdr} take approximately 60$s$ and 80$s$, respectively. The deep learning-based methods are faster than patch patch-based methods due to the training environment. Kalantari~\emph{et al.}'s method~\cite{kalantari2017deep} costs about 30$s$, which is mainly spent on the optical flow pre-processing. Wu~\emph{et al.}'s method~\cite{wu2018deep} and Yan~\emph{et al.}'s method~\cite{yan2020deep} take less inference time but their networks include a large number of parameters. Niu~\emph{et al.}'s method~\cite{niu2021hdr} and the proposed UPHDR-GAN have similar performance on the computational complexity. However, the proposed method needs fewer parameters and costs less inference time by taking the advantage of the well-designed architecture.

To better illustrate the computing efficiency of the proposed method, Fig.~\ref{fig:flops_times} shows the variation trend of the FLOPs and the inference time when selecting test images with different resolutions. The smallest image size in Fig.~\ref{fig:flops_times} is $256 \times 384$, which is labeled as $\times1$. The largest image size $\times4$ is $1024 \times 1536$. Obviously, the FLOPs and the inference time increase with the increase of image resolution. When the resolution changes from $\times3.5$ ($896 \times 1344$) to $\times4$ ($1024 \times 1536$), the FLOPs increases dramatically. If we continue to enlarge the image size, the curve of the FLOPs will have a larger slope. In order to be consistent with other methods and obtain the balance between network performance and computational complexity, we set the size of test images as $1000 \times 1500$.

\subsection{Results on Sequences Captured by Hand-held Smartphones}\label{sec:reallife}
We also conduct experiments on multi-exposure images captured by hand-held smartphones. We apply the HUAWEI Mate 10 to capture the input sequences, whose exposure time is adjusted manually. The captured scenes may have two problems: large-scale shaking and dynamic objects. To solve the first problem, we adopt the homograph registration from~\cite{guo2016joint} to achieve the background alignment. Then, the proposed architecture can handle the artifacts caused by dynamic objects. The fusion results on real-life images are shown in Fig.~\ref{fig:real_image}. The proposed method also performs well because the training dataset contains diverse scenes, including many real-life sequences captured by different devices. 

\begin{figure*}[t]
   \centering
   \includegraphics[width=0.98\textwidth]{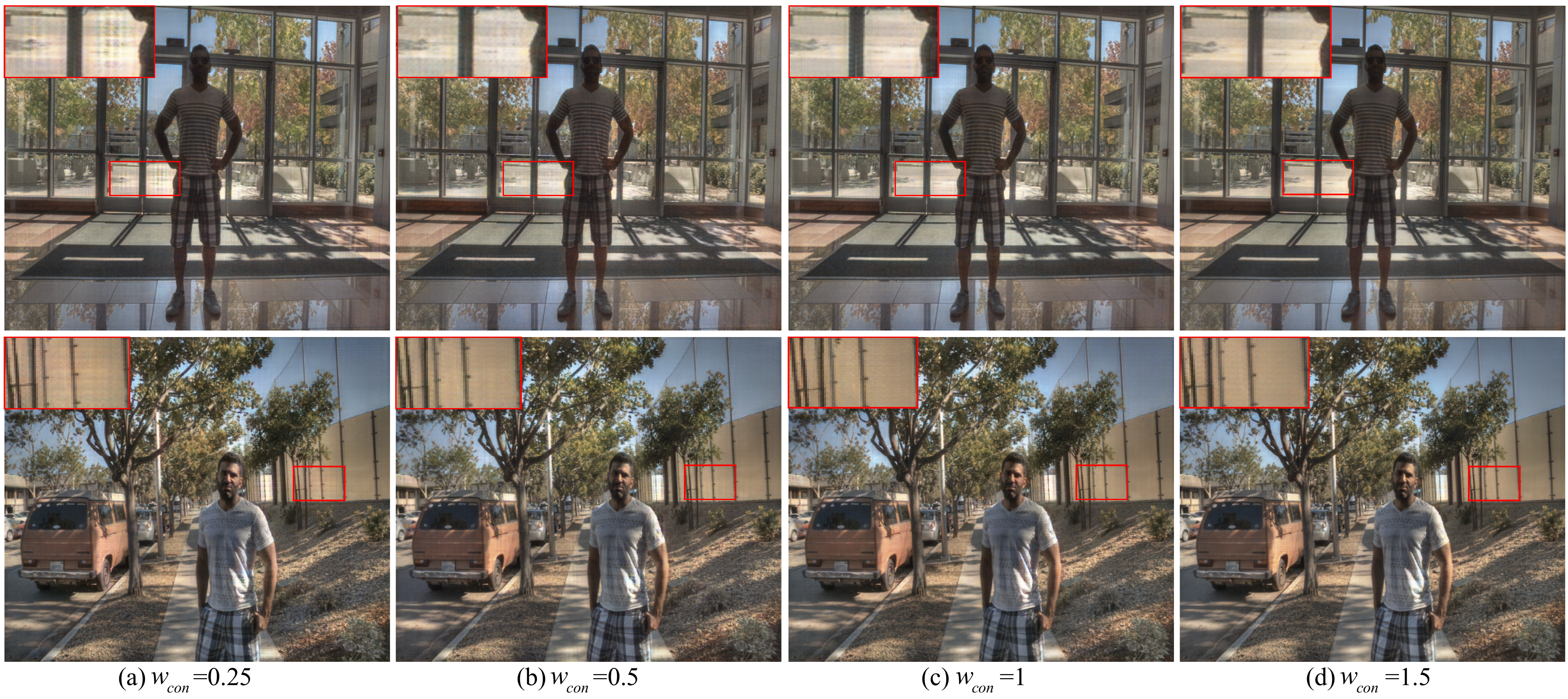}\\
   \caption{The effect of different ${w_{con}}$. We set ${w_{con}}$ to be 1.5 at the initial stage to keep a balance between HDR transformation and content preservation.}
   \label{fig:ablation_content-loss}
\end{figure*}

\begin{figure}[t]
   \centering
   \includegraphics[width=0.95\linewidth]{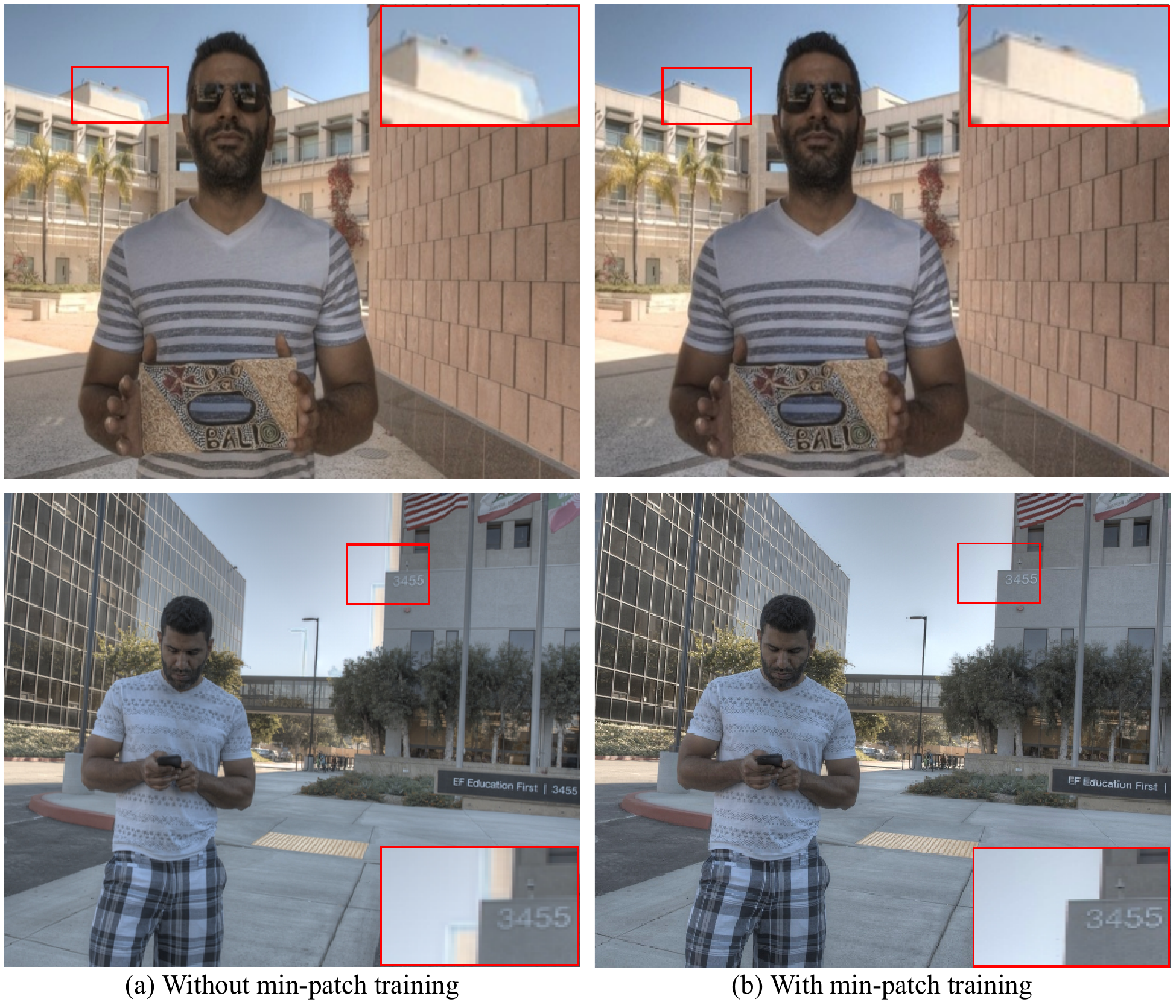}\\
   \caption{The influence of min-patch training. (a) The generated results without min-patch training. (b) The generated results with min-patch training.}
   \label{fig:ablation_min-batch}
\end{figure}

\begin{figure}[t]
  \centering
  \includegraphics[width=0.95\linewidth]{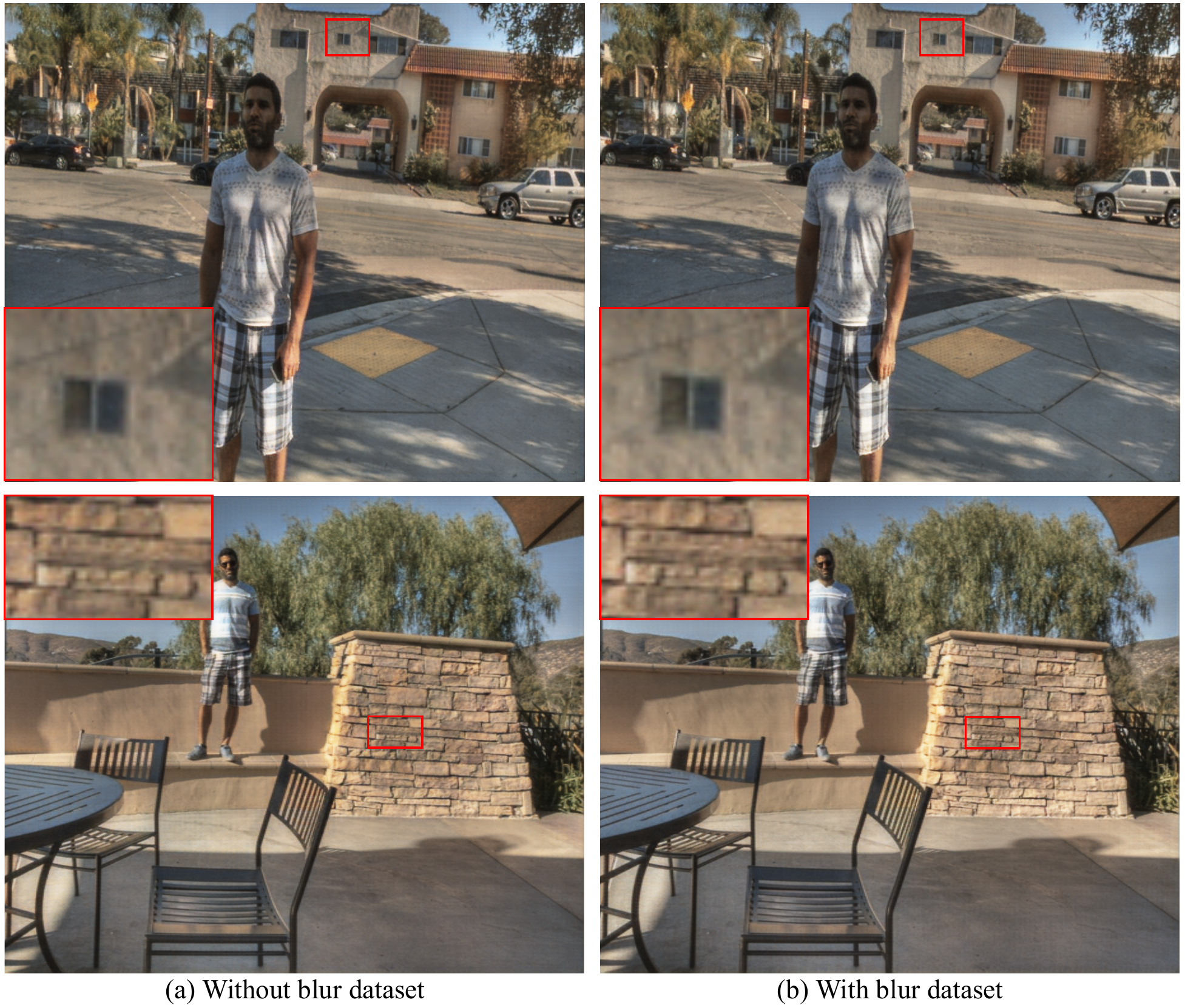}\\
  \caption{The influence of blur dataset. (a) Results without blur dataset. (b) Results with blur dataset.}
  \label{fig:ablation_blur-dataset}
\end{figure}

\subsection{Ablation Studies}\label{sec:ablation}
We conduct the ablation studies of different items in the architecture to understand the effectiveness of our designed modules.
Table~\ref{tab:ablation}
displays the ablation results of different components. First, the results from the second column to the fourth column show the importance of selecting suitable weights of the content loss. 
Second, the fifth column shows the evaluation scores when applying the MSE loss as the content loss.
Third, the results when we remove the initialization phase are listed in the sixth column in Table~\ref{tab:ablation}. Fourth, the seventh column shows the results when removing the min-patch training module. Fifth, the results without blur dataset are exhibited in the eighth column. Last, the ninth column displays the results when we merely train our network on Kalantari~\emph{et al.}'s dataset.
The results demonstrate that each component contributes to the final results. 

\subsubsection{Ablation Study of ${w_{con}}$}
We first conduct the experiments of selecting different ${w_{con}}$ to illustrate why we set the weight to 1.5. The results when we select different ${w_{con}}$ are shown in Table~\ref{tab:ablation} and Fig.~\ref{fig:ablation_content-loss}. 
From the second column to the fourth column in Table~\ref{tab:ablation}, we can conclude that unsuitable weights of the content loss apparently degrade the results, which has a consistent performance with the qualitative results in Fig.~\ref{fig:ablation_content-loss}.
Smaller ${w_{con}}$ cannot generate desired details or suffer from ghosting artifacts because they tend to learn the translation but ignore preserving the semantic content information (Fig.~\ref{fig:ablation_content-loss} (a)-(c)). 
We set ${w_{con}}$ to be 1.5 when the network in the initialization to strike a balance between unpaired domain transformation and paired semantic information preservation. 
If we continue to increase the value of ${w_{con}}$, the results will be similar to the middle-exposure LDR image because they bring more content information from the input so that the dynamic range of the result is limited.

\begin{figure*}[t]
   \centering
   \includegraphics[width=1.0\textwidth]{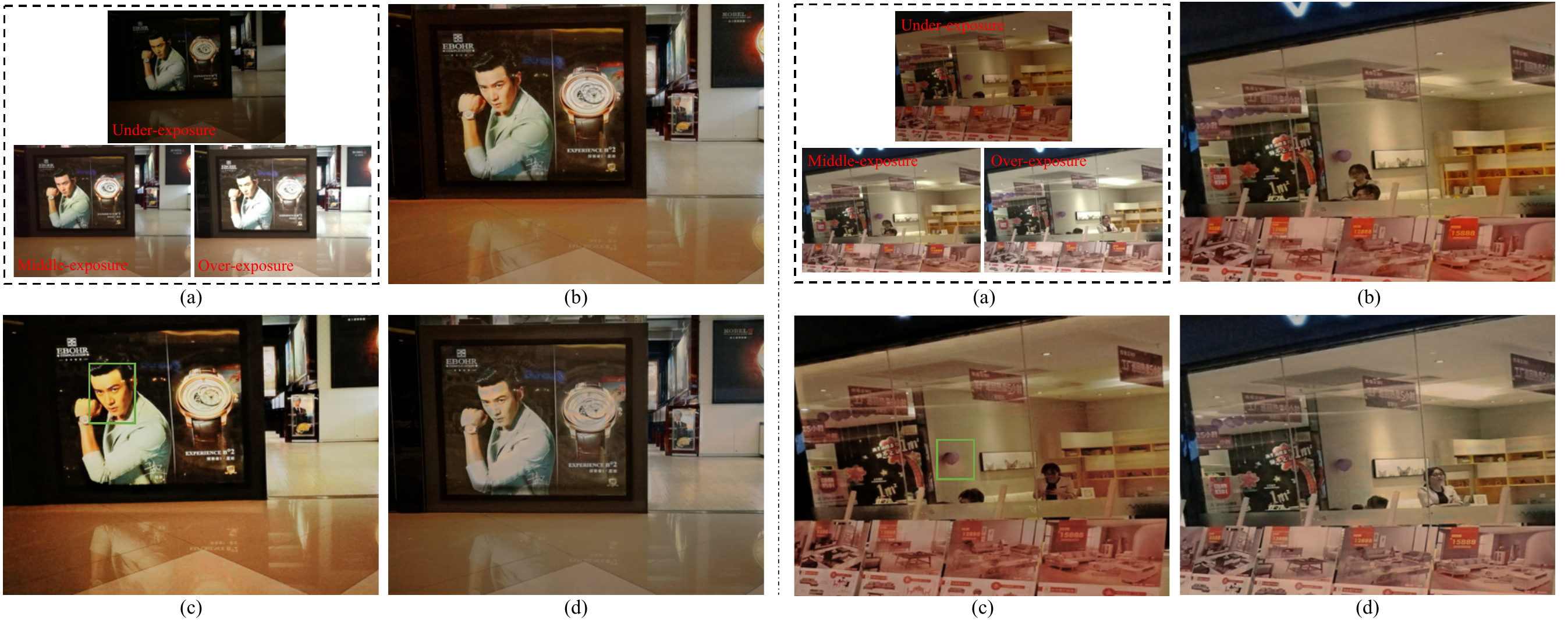}\\
   \caption{Fusion results when selecting different input images as the reference. (a) Input images. (b) Results when selecting the middle-exposure image as the reference. (c) Results when selecting the under-exposure image as the reference. (d) Results when selecting the over-exposure image as the reference.}
   \label{fig:other_middle}
\end{figure*}

\begin{figure}[!ht]
   \centering
   \includegraphics[width=1.0\linewidth]{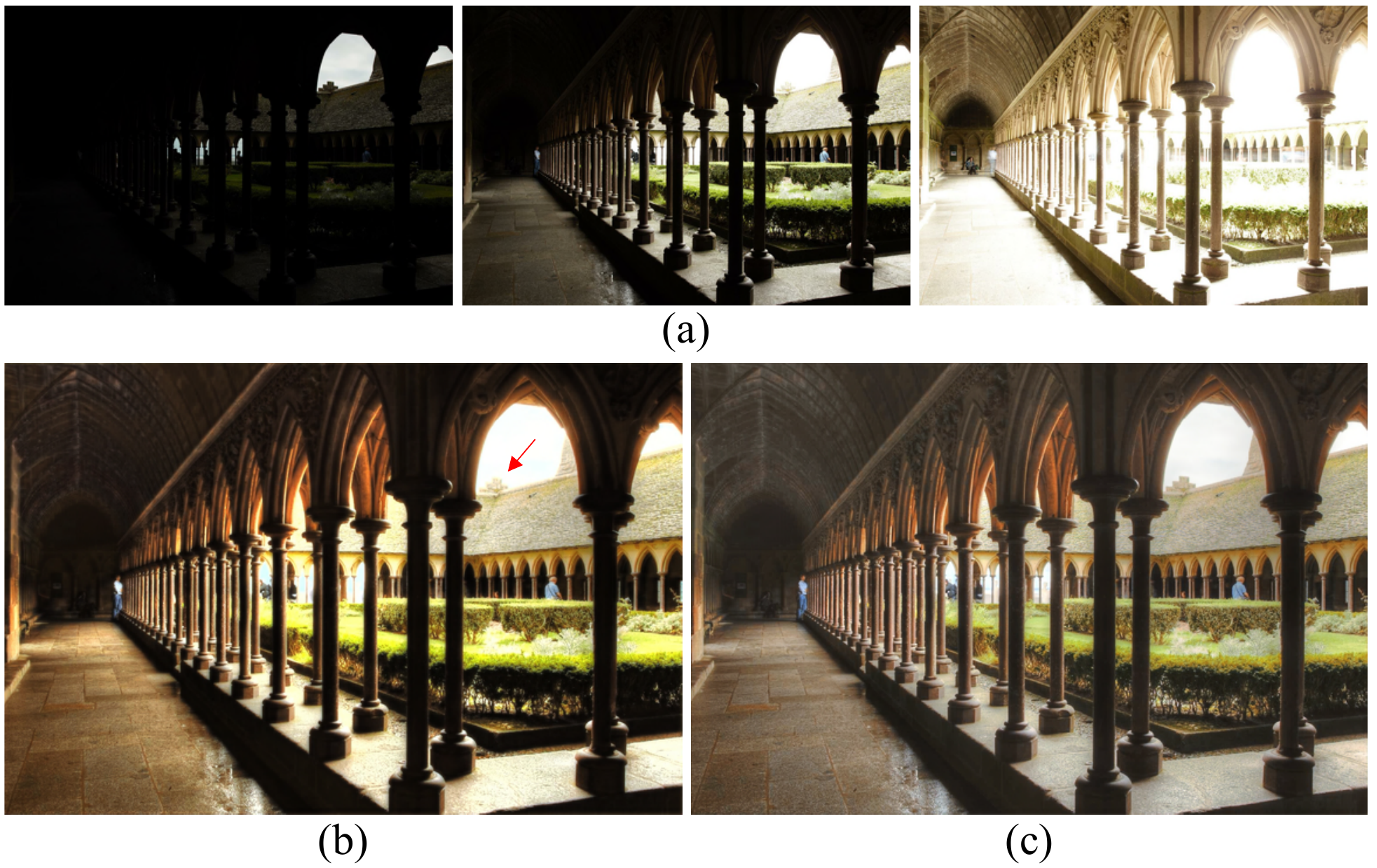}\\
   \caption{Ablation experiment of different content loss. (a) Input images. (b) Result when selecting the MSE loss as content loss. (c) Result when selecting the perceptual loss as content loss.}
   \label{fig:ablation_diffcontent}
\end{figure}

\subsubsection{Ablation Study of Min-patch Training Module}
Conventional discriminator can distinguish the real HDR images and generate HDR images. However, not all regions contribute to the discriminator optimization during training. If a small part of the generated image is so strange as to be different from the real image, it can be considered as ghosting artifact. We add the min-patch training module to detect such regions and avoid ghosting artifacts. 
The quantitative results when removing the min-patch training module in Table~\ref{tab:ablation} (the seventh column) are worse than the complete UPHDR-GAN.
Fig.~\ref{fig:ablation_min-batch} shows the effectiveness of the min-patch training module. After using the min-patch training module, UPHDR-GAN generates results with fewer artifacts (Fig.~\ref{fig:ablation_min-batch} (b)) compared to results without the min-patch training module (Fig.~\ref{fig:ablation_min-batch} (a)).

\subsubsection{Ablation Study of Blur Dataset}
Simply applying GAN loss is not sufficient for generating sharp HDR images.
Having clear edges is an important characteristic of HDR images, but common GAN loss may produce results with unclear edges. To solve the problem, we add a blur dataset $B$ as fake images to confuse the discriminator to produce images with sharp edges.
The eighth column in Table~\ref{tab:ablation} presents the quantitative results when we remove the blur dataset. Corresponding evaluation scores are lower than the final results.
Fig.~\ref{fig:ablation_blur-dataset} shows the qualitative results of without and with the blur dataset, among which the results with blur dataset (Fig.~\ref{fig:ablation_blur-dataset} (b)) have more sharp edges, such as the line shadow in the window region of the top case and the boundaries of the ceramic tiles in the bottom case.

\subsubsection{Ablation Study of Different Reference}
We also conduct the experiments when selecting different input images as the reference.
Fig.~\ref{fig:other_middle} (a) are the input images with different exposures.
Fig.~\ref{fig:other_middle} (b) are the results when selecting the middle-exposure image as the reference, while Fig.~\ref{fig:other_middle} (c) and Fig.~\ref{fig:other_middle} (d) are the results when selecting the under-exposure image and the over-exposure image as the reference, respectively. 
The two scenes in Fig.~\ref{fig:other_middle} have background misalignments between the input images. Furthermore, there is a moving person in the right case.
The proposed method can handle the misalignments and solve the moving objects well no matter which input image is chosen as the reference. For example, in the right case, when we select the under-exposure image as the reference, the semantic information of the fusion result (Fig.~\ref{fig:other_middle} (c)) is the same to the under-exposure image. The proposed method can properly handle the moving objects when fusing information from other exposure images.
However, the image quality between (b), (c) and (d) are different. 
The under-exposure image has large black regions due to the insufficient exposure time. If we select the under-exposure image as the reference, the result may suffer from color-drift (green boxes in Fig.~\ref{fig:other_middle} (c)). On the contrary, if we choose the over-exposure image as the reference, the content of over-exposed regions cannot recover well because noise can be easily introduced (Fig.~\ref{fig:other_middle} (d)). 
Obtaining information from near exposure is easy. It is challenging to acquire information from over-exposure image when the under-exposure image is selected as the reference, and vice versa.
It is reasonable that the image quality of Fig.~\ref{fig:other_middle} (c) and (d) is slightly inferior to Fig.~\ref{fig:other_middle} (b) because the target HDR domain is the collection of HDR images that correspond to the distribution of 2-nd input images. Suitable techniques to adjust the exposure are necessary to generate high-quality results when selecting the under- and over- exposure images as the reference.

\subsubsection{Ablation Study of Different Content Loss}
We show the results when applying different forms of the content loss in the fifth column of Table~\ref{tab:ablation} and Fig.~\ref{fig:ablation_diffcontent}. Our method adopts the perceptual loss (Fig.~\ref{fig:ablation_diffcontent} (c)) as the content loss to achieve high-level feature abstraction, which keeps the content information between the middle-exposure image and the generated image although the middle-exposure image and the result have different styles. Fig.~\ref{fig:ablation_diffcontent} (b) shows the result when selecting the MSE loss as the content loss, which means ${L_{con}}(G) = {\mathbb{E}_{x\sim{p_{{\rm{data}}}}(x)}} \big[(G(x) - x_2)^{2}\big]$. The MSE loss is more strict than the perceptual loss because it directly minimizes the difference between two images. As for our task, the MSE loss tends to constrain the generated image to be similar to the reference, and cannot learn the domain transformation satisfactorily. The result in Fig.~\ref{fig:ablation_diffcontent} (b) has large black regions and cannot acquire the details of under- and over-exposed regions from other exposure images (red arrow). In Table~\ref{tab:ablation}, the quantitative scores when selecting the MSE loss as content loss are also lower than the perceptual loss.

\subsection{Discussion}
Multi-exposure image fusion is a challenging topic, especially considering the image quality of generated images (related to the under- or over-exposed regions) and the ghosting artifacts (caused by the moving objects). Although we have collected a dataset that includes a variety of scenes and can satisfy recent requirements, creating a larger comprehensive dataset with more diverse scenes is helpful for the development of image fusion. Besides, as for deep learning-based methods, the number of input images is commonly fixed to three due to the network architecture. We also consider increasing the flexibility of input exposure numbers as our future work. This may be implemented by using a fully convolutional network, which is shared by different exposed images, enabling the network to process arbitrary spatial resolution and arbitrary number of exposures.

\section{Conclusion}
We have proposed a novel method to generate HDR images from multi-exposure inputs with unpaired datasets. The proposed method relaxes the constraints that deep learning-based methods need paired inputs and ground truth by introducing generative adversarial networks. The proposed method learns the translation between the input domain and the target domain and transforms the multi-inputs into an informative HDR output. 
However, generative adversarial networks obtain unclear results sometimes. We designed specific techniques to generate images with sharp edges and clear content information, including the initialization phage, the improved adversarial loss and the designed min-patch training module. Comprehensive experiments have been conducted to demonstrate the effectiveness of the proposed UPHDR-GAN.

\ifCLASSOPTIONcaptionsoff
  \newpage
\fi


\end{document}